\documentclass[aps,prb,showpacs,showkeys,preprint,eqsecnum,floatfix,superscriptaddress,nofootinbib]{revtex4-2}

\usepackage{amssymb}
\usepackage{amsmath}
\usepackage{amsbsy}
\usepackage{bm}
\usepackage{color}
\usepackage{graphicx}
\usepackage[normalem]{ulem} 
\usepackage{calrsfs}

\newcommand{\bq}{\begin{eqnarray}}
\newcommand{\eq}{\end{eqnarray}}
\newcommand{\bqn}{\begin{eqnarray*}}
\newcommand{\eqn}{\end{eqnarray*}}
\newcommand{\nn}{{\bf n}}
\newcommand{\rr}{\mathbf{r}}
\newcommand{\kk}{\mathbf{k}}

\newcommand{\LL}{{\bf L}}
\newcommand{\rrp}{{\bf r}^\prime}
\newcommand{\nablab}{\pmb{\nabla}}

\newcommand{\calp}{{\cal P}}
\newcommand{\calh}{{\cal H}}
\newcommand{\calt}{{\cal T}}
\newcommand{\calv}{{\cal V}}
\newcommand{\cala}{{\cal A}}
\newcommand{\calo}{{\cal O}}

\begin{document}
\title{Jellium at finite temperature} 

\author{Riccardo Fantoni}
\email{rfantoni@ts.infn.it}
\affiliation{Universit\`a di Trieste, Dipartimento di Fisica, strada
  Costiera 11, 34151 Grignano (Trieste), Italy}
\date{\today}

\begin{abstract}
We adopt the fixed node restricted path integral Monte Carlo method
within the ``Worm algorithm'' to simulate Wigner's Jellium model at
finite, non zero, temperatures using free-particle nodes of the
density matrix. The new element is that we incorporate the Worm
algorithm paradigm of Prokof'ev and Svistunov in the grand canonical 
ensemble in order to more
efficiently handle the fermionic exchanges. We present results for
the structure and thermodynamic properties of the ideal Fermi gas and
three points for the interacting electron gas. We treat explicitly the
case of the partially polarized electron gas. 
\end{abstract}

\keywords{Jellium,Monte Carlo simulation,finite temperature,path
  integral,worm algorithm,restricted path integral,fermions sign
  problem,structure,thermodynamic properties}

\pacs{02.70.Ss,05.10.Ln,05.30.Fk,05.70.-a,61.20.Ja,61.20.Ne}

\maketitle
\section{Introduction}
\label{sec:introduction}

The free electron gas or the {\sl Jellium} model of Wigner
\cite{Fantoni2013,Fantoni2021} is the simplest physical model for the valence
electrons in a metal \cite{Ashcroft-Mermin} (more generally it is an
essential ingredient for the study of ionic liquids (see
Ref. \cite{Hansen} Chapter 10 and 11): molten-salts, liquid-metals,
and ionic-solutions) or the plasma in the 
interior of a white dwarf \cite{Shapiro-Teukolsky}. It can be imagined
as a system of pointwise electrons of charge $e$ made 
thermodynamically stable by the presence of a uniform inert
neutralizing background of opposite charge density inside which they
move. In this work we will only be interested in the jellium in the
three dimensional Euclidean space, leaving its study in a curved
surface \cite{Fantoni03jsp,Fantoni2008,Fantoni2012,Fantoni2012b} to
later studies. 

The zero temperature, ground-state, properties of the statistical
mechanical system thus depends just on the electronic density $n$, or
the Wigner-Seitz radius $r_s=(3/4\pi n)^{1/3}/a_0$ where $a_0$ is Bohr
radius, or the Coulomb coupling parameter $\Gamma=e^2/(a_or_s)$. Free
electrons in metallic elements \cite{Ashcroft-Mermin} has $2\lesssim
r_s \lesssim 4$ whereas in the interior of a white dwarf
\cite{Shapiro-Teukolsky} $r_s\simeq 0.01$. 

The recent two decades have witnessed an impressive progress in
experiments and also in quantum Monte Carlo simulations which have
provided the field with the most accurate thermodynamic data
available. These simulations started with the work by Ceperley and
co-workers and Filinov and co-workers for jellium
\cite{Brown2013,Brown2014,Schoof2011,Schoof2015,Dornheim2015,Dornheim2016,Groth2017,Malone2016,Filinov2015},
hydrogen, hydrogen-helium mixtures and electron-hole plasmas
in the 1990s and have been improved dramatically. We recently also
applied our newly developed method to the binary fermion-boson plasma
mixture at finite temperature \cite{Fantoni2018b}, where we discussed the
thermodynamic stability of the two component mixture where the two
species are both bosons, both fermions, and one boson and one fermion. 

According to Lindhard theory of static screening, \cite{March-Tosi}
suppose we switch on an appropriately screened test charge potential 
$\delta V$ in a free electron gas. The Hartree potential $\delta
V(\rr)$ created at a distance $r$ from a static point charge of
magnitude $e$ at the origin, should be evaluated self-consistently
from the Poisson equation, 
\bq
{\bf \nabla}^2\delta V(\rr)=-4\pi e^2[\delta(\rr)+\delta n(\rr)]~~, 
\eq
where $\delta n(\rr)$ is the change in electronic density induced by
the test charge. The electron density $n(\rr)$ may be written as 
\bq
n(\rr)=2\sum_\kk |\psi_\kk(\rr)|^2~~,
\eq
where $\psi_\kk(\rr)$ are single-electron orbitals, the sum over $\kk$
is restricted to occupied orbitals ($|\kk|\leq k_F$, where $k_F$ is
the Fermi wave vector) and the factor 2 comes from the sum over spin
orientations. We must now calculate how the orbitals in
the presence of the test charge, differ from plane waves 
$\exp(i\kk\cdot\rr)$. We use for this purpose the Schr\"{o}dinger
equation, 
\bq
{\bf \nabla}^2\psi_\kk(\rr)+\left[k^2-\frac{2m}{\hbar^2}\delta V(r)\right]
\psi_\kk(\rr)=0~~,
\eq  
having imposed that the orbitals reduce to plane waves with energy 
$\hbar^2 k^2/(2m)$ at large distance \footnote{This approach (which
  lead to the Random Phase Approximation, RPA) is approximate insofar
  as the potential entering the Schr{\"o}dinger equation has been
  taken as the Hartree potential, thus neglecting exchange and
  correlation between an incoming electron and the electronic
  screening cloud.}. 

With the aforementioned boundary condition the Schr\"{o}dinger equation
may be converted into an integral equation,
\bq
\psi_\kk (\rr)=\frac{1}{\sqrt{\Omega}} e^{i\kk\cdot\rr}+\frac{2m}{\hbar^2}
\int G_{\kk}(\rr-\rrp)\delta V(\rrp)\psi_\kk(\rrp) d\rrp~~,
\eq
with $G_\kk(\rr)=-\exp(i\kk\cdot\rr)/(4\pi r)$ and $\Omega$ the volume
of the system. 

Within linear response theory we can replace $\psi_\kk(\rr)$ by
$\Omega^{-1/2}\exp(i\kk\cdot\rr)$ inside the integral. This yields 
\bq
\delta n(\rr)=-\frac{mk_F^2}{2\pi^3\hbar^2}\int j_1(2k_F|\rr-\rrp|) 
\frac{\delta V(\rrp)}{|\rr-\rrp|^2}d\rrp~~,
\eq 
with $j_1(x)$ being the first-order spherical Bessel function 
$[\sin(x)-x\cos(x)]/x^2$. Using this result in the Poisson equation
we get
\bq
{\bf \nabla}^2\delta V(r)=-4\pi e^2\delta(\rr)+\frac{2mk_F^2 e^2}
{\pi^2\hbar^2}\int j_1(2k_F|\rr-\rrp|)\frac{\delta V(\rrp)}{|\rr-\rrp|^2}
d\rrp~~,
\eq 
which is easily soluble in Fourier transform. Writing $\delta V(k)=
4\pi e^2/[k^2\varepsilon(k)]$ we find,
\bq
\varepsilon(k)=1+\frac{2mk_Fe^2}{\pi k^2\hbar^2}\left[1+\frac{k_F}{k}
\left(\frac{k^2}{4k_F^2}-1\right)\ln\left|\frac{k-2k_F}{k+2k_F}\right|
\right]~~,
\eq
which is the static dielectric function in RPA.

For $k\rightarrow 0$ this expression gives $\varepsilon(k)\rightarrow
1+k_{TF}^2/k^2$ with $k_{TF}=3\omega_p^2/v_F^2$ ($\omega_p$ being the 
plasma frequency and $v_F$ the Fermi velocity) i.e. the result of the
Thomas-Fermi theory. However $\varepsilon(k)$ has a singularity at 
$k=\pm 2k_F$, where its derivative diverges
logarithmically \footnote{The discontinuity in the momentum
  distribution across the Fermi surface introduces a singularity in
  elastic scattering processes with  momentum transfer equal to $2k_F$.}.
This singularity in $\delta V(k)$ determines, after Fourier transform,
the behavior of $\delta V(r)$ at large $r$. $\delta V(r)$ turns out to
be an oscillating function \cite{Friedel1958}
rather than a monotonically decreasing function as in the Thomas-Fermi 
theory. Indeed,
\bq
\delta V(r)=\int\frac{d\kk}{(2\pi)^3}\frac{4\pi e^2}{k^2\varepsilon(k)}
e^{i\kk\cdot\rr}=\frac{e^2}{i\pi r}\int_{-\infty}^\infty dk\frac{e^{ikr}}
{k\varepsilon(k)}~~,
\eq
and the integrand has non-analytic behavior at $q=\pm 2k_F$,
\bq
\left[\frac{1}{k\varepsilon(k)}\right]_{k\rightarrow\pm 2k_F}=
-A(k-(\pm)2k_f)\ln|k-(\pm)2k_F|+\mbox{regular terms}~~,
\eq
with $A=(k_{TF}^2/4k_F^2)/(k_{TF}^2+8k_F^2)$. Hence,
\bq
\nonumber
\delta V(r)|_{r\rightarrow \infty}&=&-\frac{Ae^2}{i\pi r}\int_{-\infty}
^\infty dk\,e^{ikr}[(k-2k_F) \ln|k-2k_F|\\
&&+(k+2k_F) \ln|k+2k_F|]=-2Ae^2\frac{\cos(2k_Fr)}{r^3}~~.
\eq
This result is based on a theorem on Fourier transforms,
\cite{Lighthill} stating that the asymptotic 
behavior of $\delta V(r)$ is determined by the low-$k$ behavior as well 
as the singularities of $\delta V(k)$. Obviously, in the present case the 
asymptotic contribution from the singularities is dominant over the 
exponential decay of Thomas-Fermi type. The result implies that the screened
ion-ion interaction in a metal has oscillatory character and ranges over 
several shells of neighbors.

Today we are able to simulate on a computer the structural and
thermodynamic properties of Jellium at finite, non zero,
temperature. This allows us to predict thermodynamic states that would
be rather difficult to obtain in nature or in the laboratory. Such as
Jellium under extreme conditions, partially polarized Jellium, etc.. In
this work we will carry on some of these path integral simulations
which make use of the Monte Carlo technique, which is the best known
method to compute a path integral. \cite{Ceperley1995} The {\sl computer
experiment} is alternative to the theoretical analytical
approximations like RPA that has been developed, during the years,
with various degrees of accuracies in different thermodynamic
conditions. Such theoretical approximations generally fall into two
categories: those which extend down from the classical regime and
those which assume some interpolation between the $T =0$ and high-$T$
regimes. From the former group we recall the Debye-H\"uckel (DH) theory
which solves for the Poisson-Boltzmann equations for the classical
one-component plasma and the quantum corrections of Hansen {\sl et
  al.} \cite{Hansen1973,Hansen1975} of the Coulomb system both with 
Wigner-Kirkwood corrections (H+WK) and without (H). Clearly these
methods do not perform well in the quantum regime below the Fermi
temperature since they lack quantum exchange. 
The Random Phase Approximation (RPA) \cite{Gupta1980,Perrot1984} is a
reasonable approximation in the low-density, high-temperature limit
(where it reduces to DH) and the low-temperature, high-density limit,
since these are both weakly interacting regimes. Its failure, however,
is most apparent in its estimation of the equilibrium, radial
distribution function $g(r)$ which becomes negative for stronger
coupling. Extensions of the RPA into intermediate densities and
temperatures have largely focused on constructing local-field
corrections (LFC) through interpolation since diagrammatic resummation
techniques often become intractable in strongly-coupled
regimes. Singwi {\sl et al.} \cite{Singwi1968} introduced one such
strategy. Tanaka and Ichimaru \cite{Tanaka1986} (TI) extended this
method to finite temperatures and provided the parameterization
of the Jellium correlation energy. This method appear to perform
marginally better than the RPA at all temperatures, though it still
fails to produce a positive-definite $g(r)$ at values of $r_s > 2$. 
A third, more recent approach introduced by Perrot and Dharma-wardana
(PDW) \cite{Perrot2000} relies on a classical mapping where the
distribution functions of a classical system at temperature $T_{cf}$,
solved for through the hypernetted-chain equation, reproduce those for
the quantum system at temperature T. In a previous work, PDW showed
such a temperature $T_q$ existed for the classical system to reproduce
the correlation energy of the quantum system at $T = 0$.
\cite{Dharma2000} To extend this work to finite temperature quantum
systems, they use the simple interpolation formula
$T_{cf}=\sqrt{T^2+T^2_q}$. This interpolation is clearly valid in the
low-$T$ limit where Fermi liquid theory gives the quadratic dependence 
of the energy on $T$. Further in the high-$T$ regime, $T$ dominates
over $T_q$ as the system becomes increasingly classical. The PDW
results matches well with the simulation results in these two
limits. It is not surprising, however, that in 
the intermediate temperature regime, where correlation effects are
greatest, the quadratic interpolation fails. A contemporary, but
similar approach by Dutta and Dufty \cite{Dutta2012} uses the same
classical mapping as PDW which relies on matching the $T = 0$ pair
correlation function instead of the correlation energy. While we
expect this to give more accurate results near $T = 0$, we would still
expect a breakdown of the assumed Fermi liquid behavior near the Fermi
temperature. Strict benchmarks have only recently been presented in
Ref. \cite{Groth2017b}. Future Jellium work will include creating a
new parameterization of the exchange-correlation energy which uses the
simulation data directly. \cite{Karasiev2014,Karasiev2019,Groth2017} In 
doing so, simulations at higher densities and both 
lower and higher temperatures may be necessary in order to complete
the interpolation between the ground-state and classical limits.  

As will be made clear in Section \ref{sec:sim}, till recently, not
even through computer experiments we were able to obtain exact numerical
results, since one had to face the so called {\sl fermions sign
  problem} which had not been solved before the advent of recent
simulation \cite{Groth2017,Dornheim2016} when it was demonstrated that
the fermion sign problem can be completely avoided and exact results
(with an error below 1\%) for the thermodynamic functions can be
obtained. In other words we were not able to extract exact results not
even numerically from a simulation for fermions, unlike for bosons or
boltzmannons. Therefore, in order to circumvent the fermion sign
problem, we will here resort to the most widely used approximation in
quantum Monte Carlo that is the {\sl restricted path integral} fixed
nodes method. \cite{Ceperley1991,Ceperley1996} But unlike previous 
studies we will implement this method upon the {\sl worm} algorithm
\cite{Prokofev1998,Boninsegni2006a} in the grand canonical ensemble.
This complements our previous study \cite{Fantoni2021} carried out in the 
canonical ensemble. In this work we will be just interested in proving 
the validity of our new numerical scheme but not his accuracy. We will
then not worry about the finite size corrections, the imaginary
thermal time discretization error, and about a stringent comparison
with previous canonical ensemble studies available in literature since 
this program has been already carried on in Ref. \cite{Fantoni2021}.

The work is organized as follows: in Section \ref{sec:model} we
describe the Jellium model from a physical point of view, in Section
\ref{sec:jft} we introduce the parameter space necessary for the
description of Jellium at finite temperature, in Section \ref{sec:sim}
we describe the simulation method, in Section \ref{sec:problem} we
outline the problem we want to solve on a computer, in Section
\ref{sec:algorithm} we presents some details of our new algorithm,
Section \ref{sec:results} is for our numerical results, and in Section 
\ref{sec:conclusions} we summarize our concluding remarks.

\section{The model}
\label{sec:model}

The {\sl Jellium} model of Wigner \cite{March-Tosi, Singwi1981,
  Ichimaru1982, Martin88} is an assembly of $N_+$ spin up pointwise
electrons and $N_-$ spin down pointwise electrons of charge $e$ moving
in a positive inert background that ensures charge neutrality. The
total number of electrons is $N=N_++N_-$ and the average particle
number density is $n=N/\Omega$, where $\Omega$ is the volume of the
electron fluid. In the volume $\Omega=L^3$ there is a uniform neutralizing
background with a charge density $\rho_b=-en$. So that the total
charge of the system is zero. The fluid polarization is then
$\xi=|N_+-N_-|/N$: $\xi=0$ in the unpolarized (paramagnetic) case and
$\xi=1$ in the fully polarized (ferromagnetic) case.

Setting lengths in units of $a=(4\pi n/3)^{-1/3}$ and energies in
Rydberg's units, $\text{Ry}=\hbar^2/2ma_0^2$, where $m$ is the electron
mass and $a_0=\hbar^2/me^2$ is the Bohr radius, the Hamiltonian of
Jellium is 
\bq 
\calh&=&-\frac{1}{r_s^2}\sum_{i=1}^N\nablab_{\rr_i}^2+V(R)~,\\ \label{p-e}
V&=&\frac{1}{r_s}\left(2\sum_{i<j}\frac{1}{|\rr_i-\rr_j|}+
\sum_{i=1}^Nr_i^2+
v_0\right)~,
\eq
where $R=(\rr_1,\rr_2,\ldots,\rr_N)$ with $\rr_i$ the
coordinate of the $i$th electron, $r_s=a/a_0$, 
and $v_0$ a constant containing the self energy of the background.
Note that the presence of the neutralizing background produces the
harmonic confinement shown in Eq. (\ref{p-e}).

The kinetic energy scales as $1/r_s^2$ and the potential
energy (particle-particle, particle-background, and
background-background interaction) scales as $1/r_s$, so for small
$r_s$ (high electronic densities), the kinetic energy dominates and
the electrons behave like an ideal gas. In the limit of large $r_s$,
the potential energy dominates and the electrons crystallize into a
Wigner crystal. \cite{Wigner1934} No liquid phase is realizable within
this model since the pair-potential has no attractive parts even
though a superconducting state \cite{Leggett1975} may still be
possible (see chapter 8.9 of Ref. \cite{Giuliani-Vignale} and
Ref. \cite{Pollock1987}).

The Jellium has been solved either by integral equation theories
in its ground-state \cite{Singwi1968} or by computer experiments in
its ground-state \cite{Ceperley1980} in the second half of last
century but more recently it has been studied at finite, non zero,
temperatures by several research groups. \cite{Brown2013,Brown2014,Schoof2011,Dornheim2015,Dornheim2016,Groth2017,Malone2016,Filinov2015} 

It was shown in Ref. \cite{Schoof2015} that the data of Brown {\sl et
  al.} \cite{Brown2013,Brown2014} are incaccurate at $r_s=1$. This
appears to be a systematic error of the fixed node method so it would
be interesting to know whether this problem
may be solved with the present method which seems a promising route to
access higher densities which was not possible in the paper by Brown
{\sl et al.}.  

\section{Jellium at finite temperature}
\label{sec:jft}

For the Jellium at finite temperature it is convenient to introduce
the {\sl electron degeneracy parameter} $\Theta=T/T_F$, where $T_F$ is
the Fermi temperature
\bq
T_F=T_D\frac{(2\pi)^2}{2[(2-\xi)\alpha_3]^{2/3}},
\eq
here $\xi$ is the polarization of the fluid that can be either
$\xi=0$, for the unpolarized case, and $\xi=1$, for the fully
polarized case, $\alpha_3=4\pi/3$, and 
\bq
T_D=\frac{n^{2/3}\hbar^2}{mk_B}=
\frac{\hbar^2}{mk_B\alpha_3^{2/3}(a_0r_s)^2},
\eq
is the degeneracy temperature, \cite{Ceperley1995} for temperatures
higher than $T_D$ quantum effects are less relevant.

The state of the fluid will then depend also upon the {\sl Coulomb
coupling parameter}, $\Gamma=e^2/(a_0r_s)k_BT$. \cite{Brown2013} So that 
\bq
\Theta=\frac{r_s}{\Gamma}\left[\frac{2(2-\xi)^{2/3}\alpha_3^{4/3}}
{(2\pi)^2}\right].
\eq 

The behavior of the internal energy of the Jellium in its ground-state
($\Theta=0$) has been determined through Diffusion Monte Carlo (DMC)
by Ceperley and Alder. \cite{Ceperley1980} Three phases of the fluid
appeared, for $r_s<75$ the stable phase is the one of the unpolarized
Jellium, for $75<r_s<100$ the one of the polarized fluid, and for
$r_s>100$ the one of the Wigner crystal. They used systems from $N=38$
to $N=246$ electrons. 

\section{The simulation} 
\label{sec:sim}

The {\sl density matrix} of a system of many fermions at temperature
$k_BT=\beta^{-1}$ can be written as an integral over all paths 
$\{R_t\}$
\bq
\rho_F(R_\beta,R_0;\beta)=\frac{1}{N!}\sum_\calp(-1)^\calp\oint_{\calp
  R_0\to R_\beta}dR_t\,\exp(-S[R_t]).
\eq
the path $R_t$ begins at $\calp R_0$ and ends at $R_\beta$ and
$\calp$ is a permutation of particles labels. For nonrelativistic
particles interacting with a potential $V(R)$ the {\sl action} of the
path, $S[R_t]$, is given by (see appendix \ref{app:bloch})
\bq
S[R_t]=\int_0^\beta dt\,\left[\frac{r_s^2}{4}\left|\frac{dR_t}{dt}\right|^2
+V(R_t)\right].
\eq 
Thermodynamic properties, such as the energy, are related to the
diagonal part of the density matrix, so that the path returns to its
starting place or to a permutation $\calp$ after a time $\beta$.

To perform Monte Carlo calculations of the integrand, one makes
imaginary time discrete with a {\sl time step} $\tau$, so that one has
a finite (and hopefully small) number of time slices and thus a
classical system of $N$ particles in $M=\beta/\tau$ time
slices; an equivalent $NM$ particle classical system of ``polymers''.
\cite{Ceperley1995} 

Note that in addition to sampling the path, the permutation is also
sampled. This is equivalent to allowing the ring polymers to connect
in different ways. This macroscopic ``percolation'' of the polymers is
directly related to superfluidity as Feynman
\cite{Feynman1953a,Feynman1953b,Feynman1953c} first showed. Any
permutation can be broken into cycles. Superfluid behavior can occur
at low temperature when the probability of exchange cycles on the
order of the system size is non-negligible. The {\sl superfluid fraction} 
can be computed in a path integral Monte Carlo calculation as
described in Ref. \cite{Pollock1987}. The same method could be used to
calculate the {\sl superconducting fraction} in Jellium at low
temperature. However, the straightforward application of those
techniques to Fermi systems means that odd permutations subtract from
the integrand. This is the ``fermions sign problem''
\cite{Ceperley1991} first noted by Feynman \cite{Feynman-Hibbs} who
after describing the path integral theory for boson superfluid $^4$He,
pointed out: 
``{\sl The} [path integral] {\sl expression for Fermi particles, such
as $^3$He, is also easily written down. However in the case of
liquid $^3$He, the effect of the potential is very hard to
evaluate quantitatively in an accurate manner. The reason for this is
that the contribution of a cycle to the sum over permutations is
either positive or negative depending whether the cycle has an odd or
an even number of atoms in its length $L$.}''

Thermodynamic properties are averages over the thermal $N-$fermions
density matrix which is defined as a thermal occupation of the exact
eigenstates $\phi_i(R)$
\bq
\rho_F(R,R';\beta)=\sum_i\phi_i^*(R)e^{-\beta E_i}\phi_i(R').
\eq
The partition function is the trace of the density matrix
\bq
Z(\beta)=e^{-\beta F}=\int dR\,\rho_F(R,R;\beta)=\sum_ie^{-\beta E_i}.
\eq
Other thermodynamic averages are obtained as
\bq
\langle\calo\rangle=Z(\beta)^{-1}\int dR dR'\,\langle R|\calo|R'\rangle
\rho_F(R',R;\beta).
\eq

Note that for any density matrix the diagonal part is always positive
\bq
\rho_F(R,R;\beta)\ge 0,
\eq
so that $Z^{-1}\rho_F(R,R;\beta)$ is a proper probability
distribution. It is the diagonal part which we need for many
observables, so that probabilistic ways of calculating those
observables are, in principle, possible.

Path integrals are constructed using the product property of density
matrices
\bq
\rho_F(R_2,R_0;\beta_1+\beta_2)=\int dR_1\,
\rho_F(R_2,R_1;\beta_2)\rho_F(R_1,R_0;\beta_1),
\eq
which holds for any sort of density matrix. If the product property is
used $M$ times we can relate the density matrix at a temperature
$\beta^{-1}$ to the density matrix at a temperature $M\beta^{-1}$. The
sequence of intermediate points $\{R_1,R_2,\ldots,R_{M-1}\}$ is the
path, and the {\sl time step} is $\tau=\beta/M$. As the time step gets
sufficiently small the Trotter theorem tells us that we can assume
that the kinetic ${\cal T}$ and potential ${\cal V}$ operator commute
so that: $e^{-\tau\calh}=e^{-\tau{\cal T}}e^{-\tau{\cal V}}$ and the
{\sl primitive approximation} for the fermions density matrix is
found. \cite{Ceperley1995} The Feynman-Kac formula for the
fermions density matrix results from taking the limit $M\to\infty$.
The price we have to pay for having an explicit expression for the
density matrix is additional integrations; all together
$3N(M-1)$. Without techniques for multidimensional integration,
nothing would have been gained by expanding the density matrix into a
path. Fortunately, simulation methods can accurately treat such
integrands. It is feasible to make $M$ rather large, say in the
hundreds or thousands, and thereby systematically reduce the time-step
error.

In addition to the internal energy and the static structure of the
Jellium one could also measure its dynamic structure, the
``superconducting fraction'', the specific heat, and the pressure.
\cite{Ceperley1995}

\subsection{Restricted Path Integral Monte Carlo}
\label{sec:rpimc}
\begin{figure}[htbp]
\begin{center}
\includegraphics[width=8cm]{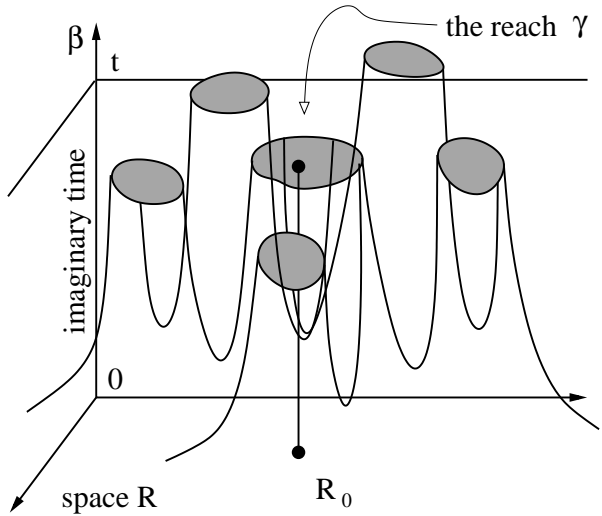}
\end{center}  
\caption{Illustration of the reach $\gamma(R_0,t)$ of the fermion
  density matrix.} 
\label{fig:the-reach}
\end{figure}

In this section we give a brief review of the restricted path integral
Monte Carlo (RPIMC) method fully
described in Refs. \cite{Ceperley1991,Ceperley1996}.
The fermion density matrix is defined by the Bloch equation which
describes its evolution in imaginary time 
\bq \label{bloch1}
\frac{\partial}{\partial\beta}\rho_F(R,R_0;\beta)&=&
-\calh\rho(R,R_0;\beta),\\ \label{bloch2}
\rho_F(R,R_0;0)&=&\cala\delta(R-R_0),
\eq
where $\beta=1/k_BT$ with $T$ the absolute temperature and $\cala$ is
the operator of antisymmetrization. The {\sl reach} of $R_0$,
$\gamma(R_0,t)$, is the set of points $\{R_t\}$ for which 
\bq
\rho_F(R_{t'},R_0;t')>0~~~0\le t'\le t,
\eq
where $\hbar t$ is the imaginary thermal time, and is illustrated in
Fig. \ref{fig:the-reach}.
Note that 
\bq
\rho_F(R_0,R_0;t)>0,
\eq
and clearly
\bq \label{rhofbc}
\rho_F(R,R_0;t)|_{R\in\partial\gamma(R_0,t)}=0.
\eq
We want to show that (\ref{rhofbc}) uniquely determines the
solution. Suppose $\delta(R,t)$ satisfies the Bloch equation
\bq
\left(\calh+\frac{\partial}{\partial t}\right)\delta(R,t)=0,
\eq
in a space-time domain $\alpha=\{t_1\le t\le t_2, R\in\Omega_t\}$
where $\Omega_t$ is the space domain at fixed imaginary thermal
time. And the two conditions
\bq \label{deltabc1}
\delta(R,t_1)&=&0,\\ \label{deltabc2}
\delta(R,t)|_{R\in\partial\Omega_t}&=&0~~~t_1\le t\le t_2,
\eq
are also satisfied. Consider
\bq \label{fun}
\int_{t_1}^{t_2}dt\,\int_{\Omega_t}dR\,e^{2V_0
  t}\delta(R,t)\left(\calh+\frac{\partial}{\partial t}\right)\delta(R,t)=0,
\eq
where $V_0$ is a lower bound for $V(R)$.

We have
\bq
\frac{\partial}{\partial t}\left[e^{2V_0 t}\delta^2(R,t)\right]=
2V_0e^{2V_0 t}\delta^2(R,t)+2e^{2V_0 t}\delta(R,t)\frac{\partial}{\partial t}
\delta(R,t). 
\eq
Since
\bq \nonumber
\int_{t_1}^{t_2}dt\int_{\Omega_t}dR\,\frac{\partial}{\partial t}
\left(\frac{e^{2V_0 t}}{2}\delta^2(R,t)\right)&=&
\int_{t_1}^{t_2}dt\,\frac{\partial}{\partial t}
\left(\frac{e^{2V_0 t}}{2}\int_{\Omega_t}dR\,\delta^2(R,t)\right)\\
&=&\frac{e^{2V_0 t_2}}{2}\int_{\Omega_{t_2}}dR\,\delta^2(R,t_2),
\eq
where in the last equality we used Eq. (\ref{deltabc1}). Then from
Eq. (\ref{fun}) follows
\bq \nonumber
&&\frac{e^{2V_0 t_2}}{2}\int_{\Omega_{t_2}}dR\,\delta^2(R,t_2)-\\
&&\int_{t_1}^{t_2}dt\,e^{2V_0 t}\int_{\Omega_t}dR\,\left[
V_0\delta^2(R,t)-\delta(R,t)\calh\delta(R,t)\right]=0.
\eq
Then using Eq. (\ref{deltabc2}) we find
\bq \nonumber
&&\frac{e^{2V_0 t_2}}{2}\int_{\Omega_{t_2}}dR\,\delta^2(R,t_2)+\\ \label{fun1}
&&\int_{t_1}^{t_2}dt\,e^{2V_0 t}\int_{\Omega_t}dR\,\left[(V(R)-V_0)\delta^2(R,t)+
\lambda\left(\nablab\delta(R,t)\right)^2\right]=0. 
\eq
With $\lambda=\hbar^2/2m$. Each term in Eq. (\ref{fun1}) is
non-negative so it must be  
\bq
\delta(R,t)=0~~~\mbox{in $\alpha$}. 
\eq
Let $\rho_1$ and $\rho_2$ be two solutions of the restricted path
problem and let $\delta=\rho_1-\rho_2$. Then
$\delta(R,t)|_{R\in\partial\gamma(R_0,t)}=0$ for $t_1\le t\le t_2$. By
taking $t_2$ to infinity and $t_1$ to zero we conclude that the
fermion density matrix is the unique solution.

Eq (\ref{fun1}) also shows that the reach $\gamma$ has the {\sl
  tiling} property. \cite{Ceperley1991} Suppose it did not. Then there
would exist a space-time domain with the density matrix non-zero
inside and from which it is only possible to reach $R_0$ or any of its
images $\calp R_0$, with $\calp$ any permutation of the particles,
crossing the nodes of the density matrix. But such a domain cannot
extend to $t=0$ because in the classical limit there are no
nodes. Then this density matrix satisfies for some $t_1>0$ the
boundary conditions (\ref{deltabc1}) and (\ref{deltabc2}) and as a
consequence it must vanish completely inside the domain contradicting
the initial hypothesis.  

We now derive the restricted path identity. Suppose $\rho_F$ is the
density matrix corresponding to some set of quantum numbers which is
obtained by using the projection operator $\cala$ on the
distinguishable particle density matrix. Then it is a solution to the
Bloch equation (\ref{bloch1}) with boundary condition (\ref{bloch2}).
Thus we have proved the {\sl Restricted Path Integral} identity
\bq \label{rpii}
\rho_F(R_\beta,R_0;\beta)=\int dR'\,\rho_F(R',R_0;0)
\oint_{R'\to R_\beta\in\gamma(R_0)}dR_t\,e^{-S[R_t]},
\eq
where the subscript means that we restrict the
path integration to paths starting at $R'$, ending at $R_\beta$ and
node-avoiding. The weight of the walk is
$\rho_F(R',R_0;0)=(N!)^{-1}\sum_\calp(-)^\calp\delta(R'-\calp
R_0)$. It is clear that the 
contribution of all the paths for a single element of the density
matrix will be of the same sign, thus solving the sign problem;
positive if $\rho_F(R',R_0;0)>0$, negative otherwise. On the
diagonal the density matrix is positive and on the path restriction
$\rho_F(R,R_0;\beta)>0$ then only even permutations are allowed
since $\rho_F(R,\calp R;\beta)=(-)^\calp\rho_F(R,R;\beta)$. It is then
possible to use a bosons calculation to get the fermions case.

Important in this argument is that the random walk is a continuous
process so we can say definitely that if sign of the density matrix
changed, it had to have crossed the nodes at some point.

The restricted path identity is one solution to Feynman's task of
rearranging terms to keep only positive contributing paths for
diagonal expectation values.

The problem we now face is that the unknown density matrix appears
both on the left-hand side and on the right-hand side of
Eq. (\ref{rpii}) since it is used to define the criterion of
node-avoiding paths. To apply the formula directly, we would somehow
have to self-consistently determine the density matrix. In practice
what we need to do is make an {\sl ansatz}, which we call $\rho_T$,
for the nodes of the density matrix needed for the restriction. The
{\sl trial density matrix}, $\rho_T$, is used to define trial nodal
cells: $\gamma_T(R_0)$. 

Then if we know the reach of the fermion density matrix we can use the
Monte Carlo method to solve the fermion problem restricting the path
integral (RPIMC) to the space-time domain where the density matrix has a
definite sign (this can be done, for example, using a trial density
matrix whose nodes approximate well the ones of the true density
matrix) and then using the antisymmetrization operator to extend 
it to the whole configuration space. This will require the complicated
task of sampling the permutation space of the $N-$particles.
\cite{Ceperley1995} Recently it has been devised an intelligent
method to perform this sampling through a new algorithm called the
{\sl worm} algorithm. \cite{Prokofev1998,Boninsegni2006a} In order to sample the
path in coordinate space one generally uses various generalizations of
the Metropolis rejection algorithm \cite{Metropolis} and the {\sl
  bisection method} \cite{Ceperley1995} in order to accomplish
multislice moves which becomes necessary as $\tau$ decreases.  

The {\sl pair-product approximation} was used by Brown {\sl et al.}
\cite{Brown2013} (see appendix \ref{app:pair-product}) to
write the many-body density matrix as a product of high-temperature
two-body density matrices. \cite{Ceperley1995} The pair Coulomb
density matrix was determined using the results of Pollock
\cite{Pollock1988} even if these could be improved using the results
of Vieillefosse. \cite{Vieillefosse1994,Vieillefosse1995} This
procedure comes with an error that scales as $\sim\tau^3/r_s^2$ where
$\tau=\beta/M$ is the {\sl time step}, with $M$ the number of
imaginary time discretizations. A more dominate form of time step error
originates from paths which cross the nodal constraint in
a time less than $\tau$. To help alleviate this effect, Brown {\sl et
  al.} \cite{Brown2013} use an image action to discourage paths from
getting too close to nodes. Additional sources of error are the finite
size one and the sampling error of the Monte Carlo algorithm
itself. For the highest density points, statistical errors are an
order of magnitude higher than time step errors. 

The results at a given temperature $T$ where obtained starting from
the density matrix in the classical limit, at small thermal times, and
using repetitively the {\sl squaring} method
\bq \label{squaring}
\rho_F(R_1,R_2;\beta)=\int dR'\,\rho_F(R_1,R';\beta/2)
\rho_F(R',R_2;\beta/2).
\eq
Time doubling is an improvement also because if we have accurate
nodes down to a temperature $T$, we can do accurate simulations down
to $T/2$. Eq. (\ref{squaring}) is clearly symmetric in $R_1$ and
$R_2$. The time doubling cannot be repeated without reintroducing the
sign problem.
 
Brown {\sl et al.} \cite{Brown2013} use $N=33$ electrons for the fully
spin polarized system and $N=66$ electrons for the unpolarized system.

\section{The Problem}
\label{sec:problem}

We need to adopt a free fermion density matrix restriction
\cite{Brown2013} for the path integral calculation from the worm
algorithm \cite{Boninsegni2006a,Boninsegni2006b} to the reach of the
reference point in the moves ending in the Z sector: remove, close,
wiggle, and displace. The worm algorithm is a particular path integral
algorithm where the permutations needs not to be sampled as they are
generated with the simulation evolution. We will use the primitive
approximation of Eq. (\ref{eq:primitive-k})-(\ref{eq:primitive-u}),
randomize the reference point time slice, restrict also the G sector,
in particular the advance and swap moves, choose the probability of
being in the G sector, $C_0$ defined in Ref. \cite{Boninsegni2006a},
as small as possible, in order not to let the worm algorithm get stuck
in the G sector when we have many time slices. Usually choosing a
smaller time step allows to use a larger $C_0$ since the path is
smoother and the restriction gives less problems in the transition
from the G to the Z sector. Or equivalently increasing the number of
time slices at fixed $C_0$ gives a larger permanence in the Z
sector. The algorithm chooses autonomously the optimal $\tau$ to be
used.  

The restriction
implementation is rather simple: we just reject the move whenever the
proposed path is such that the ideal fermion density matrix calculated
between the reference point and any of the time slices subject to 
newly generated particles positions has a negative value. Our
algorithm is described in detail in the following section.

The trial density matrix used to perform the restriction of the fixed
nodes path integral is chosen as the one of ideal fermions which is
given by 
\bq \label{ifdm}
\rho_0(R,R';t)\propto\cala \left[e^{
-\frac{(\rr_i-\rr_j')^2}{4\lambda t}}\right],
\eq
where $\lambda=\hbar^2/2m$ and $\cala$ is the antisymmetrization
operator acting on the same spin groups of particles. We expect this
approximation to be best at high temperatures and low densities when
the correlation effects are weak. Clearly in a simulation of the ideal
gas ($V=0$) this restriction returns the exact result for fermions.

We will use the primitive approximation in a grand canonical ensemble
calculation at fixed chemical potential $\mu$, volume $\Omega$, and
temperature $T$. Decreasing the chemical potential the average number
of particles diminishes. Decreasing $C_0$ the simulation spends more
time in the $Z$ sector. 

So, we will take the Bohr radius $a_0$ as units of length and energies
in Rydberg's units. In particular in the grand canonical simulation the
path integral time step $\tau~(\text{Ry}^{-1})$ will be independent
from $r_s$, unlike the simulations of Brown {\sl et al.}. \cite{Brown2013} 

The Coulomb potential is treated through the method of Natoli and
Ceperley \cite{Natoli1995} which cures its long range nature (see
Appendix \ref{app:ewald}). Even if the comparison with the direct
method by Fraser {\sl et al.} \cite{Fraser1996} gives already
reasonable results. 

We will explicitly determine the dependence of the Jellium properties
(structural and thermodynamic) on the polarization $\xi$.

\section{Our algorithm} 
\label{sec:algorithm}

Our algorithm briefly presented in the previous section is based on
the worm algorithm of Boninsegni {\sl et al.}
\cite{Boninsegni2006a,Boninsegni2006b,Fantoni2014b,Fantoni15b,Fantoni16a}. 
This algorithm uses a menu of 9 moves. 3 self-complementary: swap,
displace, and wiggle, and the 
other 6 are 3 couples of complementary moves: insert-remove,
open-close, and advance-recede. These moves act on ``worms'' with an
head {\sl Ira} and a tail {\sl Masha} in the $\beta-$periodic
imaginary thermal time, which can swap a portion of
their bodies (swap move), can move forward and backward
(advance-recede moves), can be subdivided in two
or joined into a bigger one (open-close moves), and can be born or die
(insert-remove moves) since we are working in the grand-canonical
ensemble. The configuration space of the worms is called the G
sector. When the worms recombine to 
form a closed path we eneter the so called Z sector and the path can
translate in space (displace move) and can propagate in space through
the bisection algorithm (wiggle move) carefully explained in
Ref. \cite{Ceperley1995}.  

\begin{figure}[htbp]
\begin{center}
\includegraphics[width=10cm]{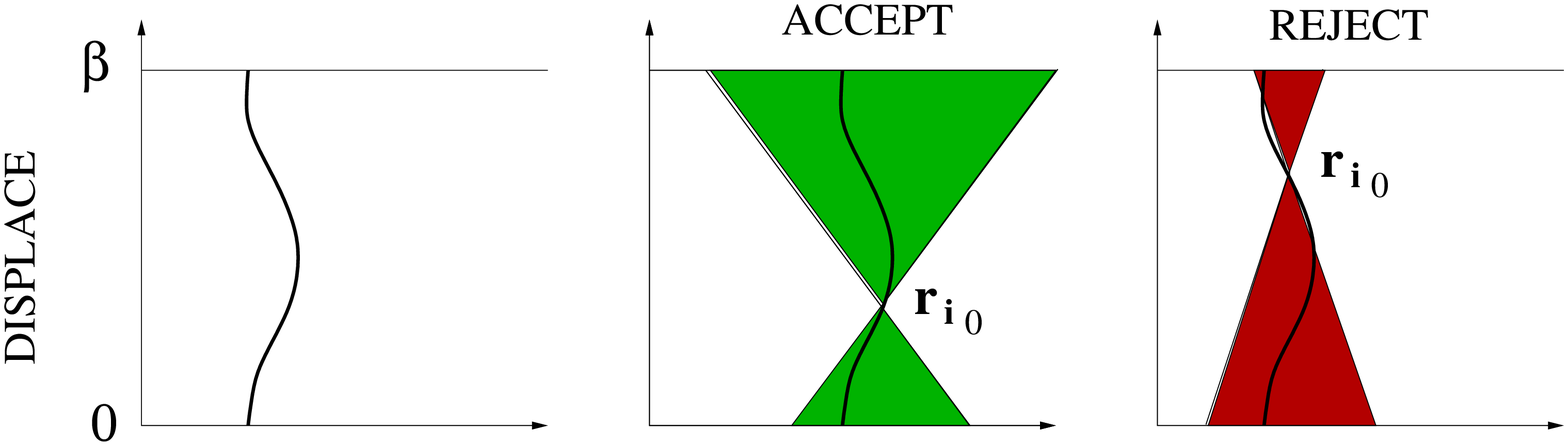}\\
\includegraphics[width=10cm]{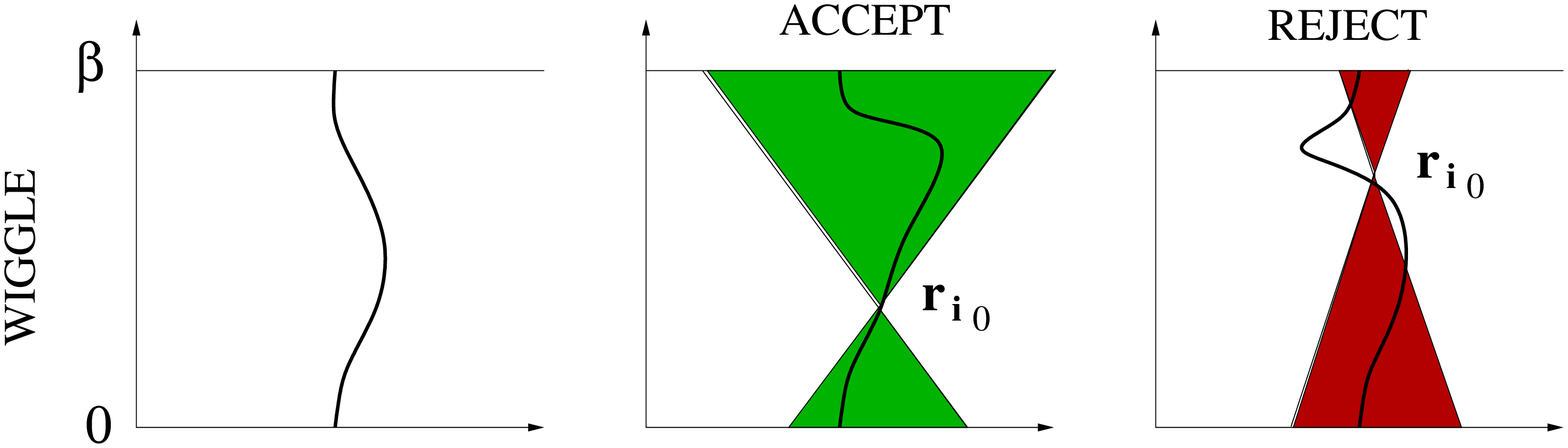}\\
\includegraphics[width=10cm]{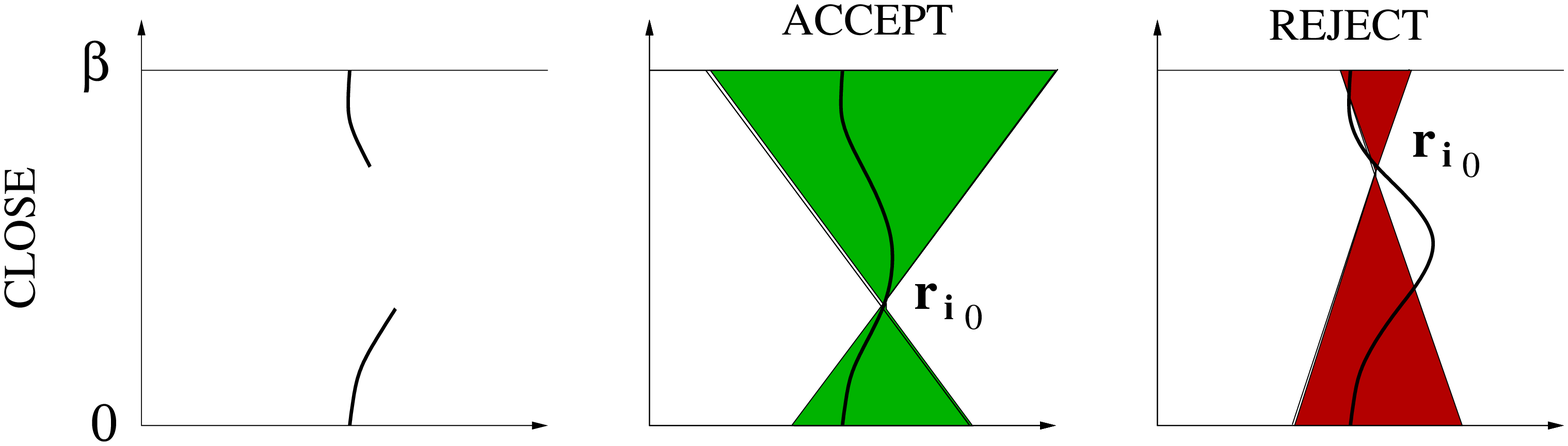}
\end{center}  
\caption{(color online) Illustration of the rejection algorithm within
  the worm algorithm. The bold line represents schematically the
  closed path or the open worm, of a single electron. In the most
  general case these will wind through the beta periodic imaginary
  thermal time circle, but this is not shown in the illustration. The
  reference point is ${\rr_i}_0$ and the microscopic reach is
  represented schematically as the shaded doubly cylindrical
  region. In general the reach will be a complicated region of
  space-time as pointed out in Fig. \ref{fig:the-reach} for the
  macroscopic reach. Only the three moves: displace (Z$\to$Z), wiggle
  (Z$\to$Z), and close (G$\to$Z) are shown. On the leftt we have the
  starting configuration and on the right we show two different
  actions of each move, one accepted and one rejected.}  
\label{fig:alg}
\end{figure}

In order to reach a restricted path integral we restrict the moves
that end in the Z sector, that is: displace, wiggle, close, and remove.
This is pictorially shown in Fig. \ref{fig:alg} for the first three
moves. It is important to stress the fact that we choose the reference
point time slice randomly (i.e. we choose an integer rantom number
between 1 and $M$, say $m$, and the reference point will then be
$R_0=R_{m\tau}$), before each move, to increse the acceptances in 
the restrictions. This is allowed because we are free to perform a
translation in the $\beta-$periodic imaginary thermal time. The
reaches of different reference points will in general be different. In
the figure the reach is schematically represented as a double cone.
In order to increase the acceptances in the restrictions we also
restricted some moves in the G sector: swap and advance. 

In order to
implement the restriction we reject the move whenever the proposed
path is such that the ideal fermion density matrix of Eq. (\ref{ifdm})
calculated between the reference point and any of the time slices
subject to newly generated particles positions has a negative
value. That is, whenever the path ends up in a region not belonging to
the reach of the reference point as shown in Fig. \ref{fig:alg}. The
restriction of the G sector moves acts in the same way but on worms
rather then on closed paths. When calculating diagonal properties we
consider only the density matrix at the reference point.

Since the averages are only taken during the permanence in the Z
sector it is fundamental to restrict the moves that end in the Z
sector. Since these are the ones that have an influence on the
measures of the various estimators during the run. If we enter the Z
sector in such a way that we are out of the reach of the reference
point the algorithm will continue wandering in the G sector till a
door to the Z sector opens up. The code without restrictions gives the
bosonic calculation so we are free to restrict also the G sector in
order to increase the acceptances of the Z sector. 

For each move we can decide the frequency of the move and the maximum
number of time slices it operates on, apart from the displace move
where instead of the maximum number of time slices we can decide the
maximum extent of the spatial translation displacement. It is well
known that Monte Carlo algorithms works better as long as we have a
longer moves menu, unless of course one violates detailed balance. So
the worm algorithm is very efficient in exploring all the electrons
path configuration with all the necessary exchanges.

\section{Results} 
\label{sec:results}

\begin{figure}[htbp]
\begin{center}
\includegraphics[width=10cm]{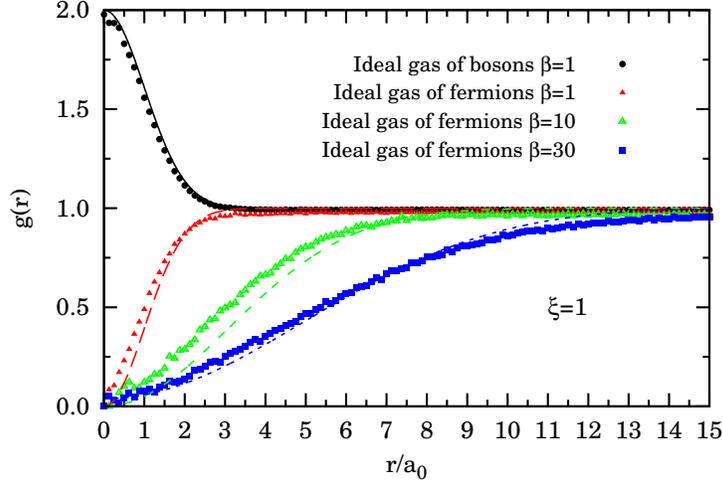}
\end{center}  
\caption{(color online) The radial distribution function for an ideal
  gas of bosons 
  at one inverse temperature ($\beta=1~\text{Ry}^{-1}$) and an ideal
  gas of fermions at three inverse temperatures
  ($\beta=1~\text{Ry}^{-1},10~\text{Ry}^{-1},30~\text{Ry}^{-1}$). We
  simulate fully polarized ($\xi=1$) particles. The
  exact analytical results are shown as guiding lines and were derived
  from the work of Bosse {\sl et al.}. \cite{Bosse2011}}  
\label{fig:ideal}
\end{figure}

In order to test the validity of the restriction procedure we first
simulated a system of free ($V=0$) particles without the
restriction (bosons) and with the restriction (fermions). The result
for the radial distribution function is shown in Fig. \ref{fig:ideal}.
The small discrepancy with the analytic result of Bosse {\sl et al.}
\cite{Bosse2011} is due to the finite size effect. The average number
of particles in the simulation for the bosons being around 107 and for
the fermions 46 for $\beta=1~\text{Ry}^{-1}$, 27 for
$\beta=10~\text{Ry}^{-1}$, and 21 for $\beta=30~\text{Ry}^{-1}$. For
the free particles we do not have any source of error coming from the
imaginary time discretization. Since we were not interested in a
quantitative accurate analysis we chose the simulations at smaller
temperatures shorter. The volume was kept fixed at
$\Omega=1.25\times 10^5a_0^3$ corresponding to a half box side of
$L/2=25a_0$. We used 20 time slices for the boson case and 80 for
the fermion cases. 
In these simulations we find good agreement with the exact analytic
results also for the internal energy per particle (kinetic and
potential) and for the pressure.

Then we simulated the Jellium using for the potential energy, $V$, the
image potential, $V_I$, of Eq. (\ref{eq:ewald}) where we chose the
short and long range splitting, necessary for the bare Coulomb potential
$v(r)=2~\text{Ry}/r$, using the optimized method of Natoli and
Ceperley \cite{Natoli1995} with an eight-order polynomial for the
radial interpolation. In the long range part we keep up to 128 Fourier
components.

In Table \ref{tab:tq} we present our results for various thermodynamic
quantities. Our results
cannot be directly compared with the ones of Brown {\sl et al.}
\cite{Brown2013} since we are running at fixed chemical potential but
we believe that we are able to extend their results at higher density
$r_s<1$. Benchmark data can be found in
Refs. \cite{Dornheim2016b,Groth2016}. We leave a careful comparison in 
a subsequent work.  
\begin{table}[htbp]
\caption{Thermodynamic results in our simulations:
  $\beta~(\text{Ry}^{-1})$ inverse temperature, $e_k~(\text{Ry})$ 
  kinetic energy per particle, $e_p~(\text{Ry})$ potential energy per
  particle, $P~(\text{Ry}/a_0^3)$ pressure.}
\label{tab:tq}
\begin{ruledtabular}
\begin{tabular}{lllllllllll}
$M$ & $\xi$ & $\overline{N}$ & $L$ & $\beta$ & $r_s$ & $\Theta$ & $\Gamma$ & $e_k$ & $e_p$ & $P$  \\
\hline
60 & 1 & 35.35(4) & 5  & 0.04 & 0.945 & 3.819 & 0.085  & 31.5(5) &
-0.736(3) & 5.7(1) \\
80 & 0.154 & 57.0(2) & 50 & 4 & 8.060 & 4.180 & 0.993 & 0.365(8) &
-0.0921(4) & 5.2(2)$\times 10^{-5}$ \\
680& 1 & 30.15(3) & 50 & 68   & 9.966 & 0.250 & 13.647 & 0.016(1) &
-0.12198(5) & $\approx$0 
\end{tabular}
\end{ruledtabular}
\end{table}

\begin{figure}[htbp]
\begin{center}
\includegraphics[width=10cm]{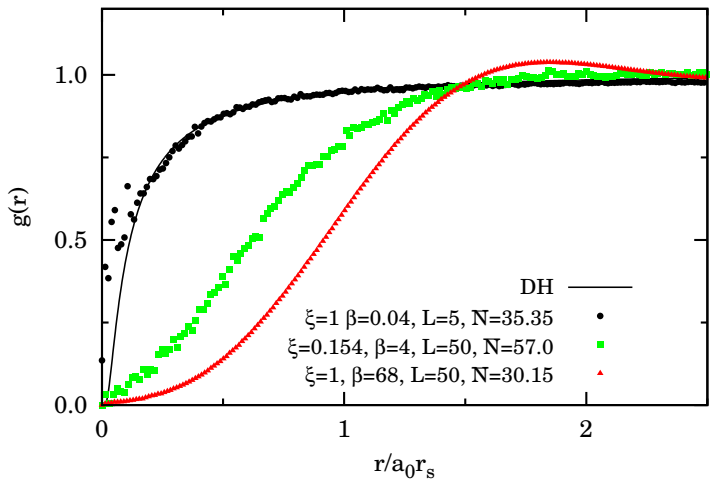}
\end{center}  
\caption{(color online) The radial distribution function for Jellium
  in the states 
  of Table \ref{tab:tq}. Also shown is the DH result for the highest
  temperature state,
  $g_{DH}(r)=\exp\left[-\frac{\Gamma}{r}\exp\left(-\sqrt{3\Gamma}r
  \right)\right]$.}
\label{fig:jellium}
\end{figure}

In Fig. \ref{fig:jellium} we show our results for the radial
distribution function for the states of Table \ref{tab:tq}.
%

\section{Conclusions} 
\label{sec:conclusions}

We have successfully implemented the ideal fermion density matrix
restriction on the path integral worm algorithm which is able to
generate the necessary permutations during the simulation evolution
without the need of their explicit sampling. This allowed us to
reach the fermionic finite temperature properties of a given fluid of
particles interacting through a pair potential. We worked in the grand
canonical ensemble and applied our method to the Jellium fluid of
Wigner. Even if our results cannot be directly compared with the
previous canonical calculation of Brown {\sl et al.} \cite{Brown2013}
(this program was already carried out in our previous work 
\cite{Fantoni2021}) we believe that they complement them with the access 
to the high density regime and with the treatment of the general 
polarization case. In this preliminary paper we just address the validity 
of our method, its accuracy will be treated in a forthcoming work.

The relevance of our study relies in the fact that our simulation
method is different from both the method of Ceperley {\sl et al.} 
\cite{Brown2013,Brown2014} who uses the fixed nodes approximation in
the canonical ensemble and explicitly samples the necessary
permutations, and from the one of Bonitz {\sl et
  al.} \cite{Schoof2011,Dornheim2015,Dornheim2016,Groth2017} who
combine configuration path 
integral Monte Carlo and permutation blocking path integral Monte
Carlo. Our method is also different from others quantum Monte Carlo
methods like the one of Malone {\sl et al.} \cite{Malone2016} that
agrees well with the one of Bonitz at high densities and the direct
path integral Monte Carlo one of Filinov {\sl et al.}
\cite{Filinov2015} that agrees well with Brown at low density and
moderate temperature. So our new algorithm adds to the ones already used
in the quest for an optimal way to calculate the properties of the
fascinating Wigner's Jellium model at finite temperatures.

We obtained results for both the structure, the radial
distribution function, and various thermodynamic quantities. 

We intend to adopt this method to simulate Jellium in a curved surface
\cite{Fantoni03jsp,Fantoni2008,Fantoni2012,Fantoni2012b}
in the near future. For example the Jellium on the surface of a sphere
with a Dirac magnetic monopole at the center could be used to study the
quantum Hall effect \cite{Melik2001}. We already successfully applied the 
present method to Jellium on the surface of a sphere \cite{Fantoni2018} 
and to two component boson-fermion plasma on a plane \cite{Fantoni2018b}.

\appendix
\section{The primitive action}
\label{app:bloch}

In this appendix we give a brief review of the derivation of the
primitive approximation given in Ref. \cite{Ceperley1995}.
Suppose the Hamiltonian is split into two pieces $\calh=\calt+\calv$,
where $\calt$ and $\calv$ are the kinetic and potential
operators. Recall the exact Baker-Campbell-Hausdorff formula to
expand $\exp(-\tau\calh)$ into the product
$\exp(-\tau\calt)\exp(-\tau\calv)$. As $\tau\to 0$ the commutator
terms which are of order higher than $\tau^2$ become smaller than the
other terms and thus can be neglected. This is known as the {\sl
  primitive approximation}
\bq
e^{-\tau(\calt+\calv)}\approx e^{-\tau\calt}e^{-\tau\calv}.
\eq
hence we can approximate the exact density matrix by product of the
density matrices for $\calt$ and $\calv$ alone. One might worry that
this would lead to an error as $M\to\infty$, with small errors
building up to a finite error. According to the Trotter
\cite{Trotter1959} formula, one does not have to worry
\bq
e^{-\beta(\calt+\calv)}=\lim_{M\to\infty}\left[e^{-\tau\calt}e^{-\tau\calv}\right]^M.
\eq
The Trotter formula holds if the three operators $\calt$, $\calv$, and
$\calt+\calv$ are self-adjoint and make sense separately, for example,
if their spectrum is bounded below. \cite{Simon1979} This is the case
for the Hamiltonian describing Jellium.

Let us now write the primitive approximation in position space
\bq
\rho(R_0,R_2;\tau)\approx\int dR_1\langle
R_0|e^{-\tau\calt}|R_1\rangle\langle R_1|e^{-\tau\calv}|R_2\rangle,
\eq
and evaluate the kinetic and potential density matrices. Since the
potential operator is diagonal in the position representation, its
matrix elements are trivial
\bq \label{primitive-v}
\langle R_1|e^{-\tau\calv}|R_2\rangle=e^{-\tau V(R_1)}\delta(R_2-R_1).
\eq

The kinetic matrix can be evaluated using the eigenfunction expansion
of $\calt$. Consider, for example, the case of distinguishable
particles in a cube of side $L$ with periodic boundary
conditions. Then the exact eigenfunctions and eigenvalues of $\calt$
are $L^{-3N/2}e^{iK_\nn R}$ and $\lambda K_\nn^2$, with
$K_\nn=2\pi\nn/L$ and $\nn$ a $3N$-dimensional integer vector. We are
using here dimensional units. Then 
\bq \label{primitive-t-1}
\langle R_0|e^{-\tau\calt}|R_1\rangle&=&\sum_\nn L^{-3N}e^{-\tau\lambda
  K_\nn^2}e^{-iK_\nn(R_0-R_1)}\\ \label{primitive-t-2}
&=&(4\pi\lambda\tau)^{-3N/2}
\exp\left[-\frac{(R_0-R_1)^2}{4\lambda\tau}\right],
\eq
where $\lambda=\hbar^2/2m$. Eq. (\ref{primitive-t-2}) is obtained by
approximating the sum by an integral. This is appropriate only if the
thermal wavelength of one step is much less than the size of the box,
$\lambda\tau\ll L^2$. In some special situations this condition could
be violated, in which case one should use Eq. (\ref{primitive-t-1}) or
add periodic ``images'' to Eq. (\ref{primitive-t-2}). The exact
kinetic density matrix in periodic boundary conditions is a theta
function, $\prod_{i=1}^{3N}\theta_3(z_i,q)$, where
$z_i=\pi(R_0^i-R_1^i)/L$, $R^i$ is the $i$th component of the $3N$
dimensional vector $R$, and $q=e^{-\lambda\tau(2\pi/L)^2}$ (see chapter
16 of Ref. \cite{Abramowitz}). Errors from ignoring the boundary
conditions are $O(q)$, exponentially small at large $M$.  

A {\sl link} $m$ is a pair of time slices $(R_{m-1},R_m)$ separated by
a {\sl time step} $\tau=\beta/M$. The {\sl action} $S^m$ of a link is
defined as minus the logarithm of the exact density matrix. Then the
exact path-integral expression becomes
\bq
\rho(R_0,R_M;\beta)=\int dR_1\ldots dR_{M-1}\,
\exp\left[-\sum_{m=1}^M S^m\right],
\eq 
It is convenient to separate out the {\sl kinetic action} from the
rest of the action. The exact kinetic action for link $m$ will be
denoted $K^m$ 
\bq \label{eq:primitive-k}
K^m=\frac{3N}{2}\ln(4\pi\lambda\tau)+\frac{(R_{m-1}-R_m)^2}{4\lambda\tau},
\eq
The {\sl inter-action} is then defined as what is left
\bq
U^m=U(R_{m-1},R_m;\tau)=S^m-K^m.
\eq
In the primitive approximation the inter-action is 
\bq \label{eq:primitive-u}
U^m_1=\frac{\tau}{2}[V(R_{m-1})+V(R_m)],
\eq
where we have symmetrized $U^m_1$ with respect to $R_{m-1}$ and $R_m$,
since one knows that the exact density matrix is symmetric and thus
the symmetrized form is more accurate. 

A capital letter $U$ refers to the total link inter-action. One should
not think of the exact $U$ as being strictly the potential
action. That is true for the primitive action but, in general, is only
correct in the small-$\tau$ limit. The exact $U$ also contains kinetic
contributions of higher order in $\tau$. If a subscript is present on
the inter-action, it indicates the order of approximation; the
primitive approximation is only correct to order $\tau$. No subscript
implies the exact inter-action.

The {\sl residual energy} of an approximate density matrix is defined
as 
\bq \label{residual-energy}
E_A(R,R';t)=\frac{1}{\rho_A(R,R';t)}\left[\calh+
\frac{\partial}{\partial t}\right]\rho_A(R,R';t).
\eq
The residual energy for an exact density matrix vanishes; it is a
local measure of the error of an approximate density matrix. The
Hamiltonian $\calh$ is a function of $R$; thus the residual energy is
not symmetric in $R$ and $R'$. 

It is useful to write the residual energy as a function of the
inter-action. We find
\bq \label{residual-energy-b}
E_A(R,R';t)=V(R)-\frac{\partial U_A}{\partial t}-
\frac{(R-R')\cdot \nabla U_A}{t}+\lambda\nabla^2U_A-
\lambda\left(\nabla U_A\right)^2.
\eq
The terms on the right hand side are ordered in powers of $\tau$,
keeping in mind that $U(R)$ is of order $\tau$, and $|R-R'|$ is of
order $\tau^{1/2}$. One obtains the primitive action by setting the
residual energy to zero and dropping the last three terms on the right
hand side. 

The residual energy of the primitive approximation is
\bq
E_1(R,R';t)=\frac{1}{2}\left[V(R)-V(R')\right]-
\frac{1}{2}(R-R')\cdot\nabla V+
\frac{\lambda t}{2}\nabla^2V-
\frac{\lambda t^2}{4}\left(\nabla V\right)^2.
\eq
With a leading error of $\sim \lambda\tau^2$.

\section{The pair-product action}
\label{app:pair-product}

An often useful method to determine the many-body action is to use the
exact action for two electrons. \cite{Barker1979} To justify this
approach, first assume that the potential energy can be broken into a
pairwise sum of terms
\bq
V(R)=\sum_{i<j}v(|\rr_i-\rr_j|),
\eq
with $|\rr_i-\rr_j|=r_{ij}$.
Next, apply the Feynman-Kac formula for the inter-action
\bq
e^{-U(R_0,R_F;\tau)}=\left\langle\exp\left[-\int_0^\tau dt\,V(R(t))
\right]\right\rangle_\text{RW},
\eq
where the notation $\langle\ldots\rangle_\text{RW}$ means the average
over all Gaussian random walks from $R_0$ to $R_F$ in a ``time''
$\tau$. So that 
\bq
e^{-U(R_0,R_F;\tau)}&=&\left\langle\exp\left[-
\int_0^\tau dt\,\sum_{i<j}v(r_{ij}(t))\right]\right\rangle_\text{RW}\\
&=&\left\langle\prod_{i<j}\exp\left[-\int_0^\tau dt\,v(r_{ij}(t))\right]
\right\rangle_\text{RW}\\
&\approx&\prod_{i<j}\left\langle\exp\left[-\int_0^\tau dt\,v(r_{ij}(t))\right]
\right\rangle_\text{RW}\\
&=&\prod_{i<j}\exp\left[-u_2(r_{ij},r'_{ij};\tau)\right]\\
&=&\exp\left[-\sum_{i<j}u_2(r_{ij},r'_{ij};\tau)\right]
=e^{-U_2(R_0,R_F;\tau)},
\eq
where $U_2$ is the {\sl pair-product} action and $u_2$ is the exact
action for a pair of electrons. At low temperatures the pair action
approaches the solution of the two particle wave equation. The result
is the pair-product or Jastrow ground-state wave function, which is
the ubiquitous choice for a correlated wave function because it does
such a good job of describing most ground-state correlations.

The residual energy (see Eq. (\ref{residual-energy})) for the
pair-product action is less singular than for other forms. We have
that 
\bq
u_2(r_{ij},r_{ij}';\tau)=-\ln\left\langle\exp\left(
-\int_0^\tau dt\,v(r_{ij}(t))\right)\right\rangle_\text{RW},
\eq
is of order $\tau^2$ since the two body problem can be factorized into
a center-of-mass term and a term that is a function of the relative
coordinates. Moreover we must have
\bq
\frac{\partial u_2}{\partial\tau}=v(r_{ij}(\tau)),
\eq
so that 
\bq
\frac{\partial U_2}{\partial\tau}=V(R(\tau)),
\eq
which tells that only the last three terms on the right hand side of
Eq. (\ref{residual-energy-b}) contribute to the residual energy. We
also have  
\bq
\nabla U_2=\sum_i\sum_{i\ne j}\nabla_iu_2(r_{ij},r_{ij}';\tau),
\eq
where the indices run over the particles. So the leading error of the
pair-product action is $\sim \lambda\tau^3$. 

\section{Long-range potentials with the Ewald image technique}
\label{app:ewald}

Suppose the bare potential in infinite $d$ dimensional space is
$v(r)$. Let us define the Fourier transform by 
\bq
\tilde{v}_\kk=\int_{-\infty}^\infty d^d\rr\,e^{-i\kk\cdot\rr}v(r).
\eq
Then its inverse is
\bq
v(r)=\int_{-\infty}^\infty \frac{d^d\kk}{(2\pi)^d}\,e^{i\kk\cdot\rr}
\tilde{v}_\kk.
\eq

Now let us find the energy of a single particle interacting with an
infinite rectangular lattice of another particle a distance $\rr$
away. To make it converge we also add a uniform background of the same
density ($\Omega=$volume) of opposite charge. Thus the ``image
pair-potential'' is equal to
\bq
v_I(r)=\sum_{\LL}v(\rr+\LL)-\tilde{v}_0/\Omega.
\eq 
The $\LL$ sum is over the Bravais lattice of the simulation cell
$\LL=(m_xL_x,m_yL_y,\ldots)$ where $m_x,m_y,\ldots$ range over all
positive and negative integers. Converting this to $k-$space and using
the Poisson sum formula we get
\bq
v_I(r)=\frac{1}{\Omega}\sum_{\kk}'\tilde{v}_\kk e^{i\kk\cdot\rr},
\eq
where the prime indicates that we omit the $\kk=0$ term; it cancels
out with the background. The $\kk-$sum is over reciprocal lattice
vectors of the simulation box $\kk=(2\pi n_x/L_x,2\pi
n_y/L_y,\ldots)$.

Because both sums are so poorly convergent, we make the division into
$k-$space and $r-$space; taking the long-range part into $k-$space. We
write 
\bq
v(r)=v_s(r)+v_l(r)~,
\eq
where the optimal splitting is discussed in the work by Natoli and
Ceperley. \cite{Natoli1995} Since Fourier transform is linear, we can
also write
\bq
\tilde{v}_\kk=\tilde{v}_{s\kk}+\tilde{v}_{l\kk}.
\eq
Then the image pair-potential is written as
\bq
v_I(r)=\sum_\LL
v_s(|\rr+\LL|)+\frac{1}{\Omega}\sum_\kk\tilde{v}_{l\kk}
e^{i\kk\cdot\rr}-\frac{1}{\Omega}\tilde{v}_0.
\eq

Now let us work with $N$ particles of charge $q_i$ in a periodic box
and let us compute the total potential energy of the unit
cell. Particles $i$ and $j$ are assumed to interact with a
pair-potential $q_iq_jv(r_{ij})$. The image potential energy for the
$N-$particle system is   
\bq
V_I=\sum_{i<j}q_iq_jv_I(r_{ij})+\sum_{i}q_i^2v_M,
\eq
where $v_M$ is the interaction of a particle with its own images; it
is a Madelung constant for particle $i$ interacting with the perfect
lattice of the simulation cell. If this term were not present,
particle $i$ would only see $N-1$ particles in the surrounding cells
instead of $N$. We can find its value by considering the limit as two
particles get close together with the image pair-potential. Hence
\bq
v_M=\frac{1}{2}\lim_{r\to 0}[v_I(r)-v(r)].
\eq
Now we substitute the split up image pair-potential and collect all
the terms together
\bq \nonumber
V_I&=&\sum_{i<j}\sum_{\LL}q_iq_jv_s(|\rr_{ij}+\LL|)
+\frac{1}{\Omega}\sum_\kk'\tilde{v}_{l\kk}\sum_{i<j}q_iq_j
e^{i\kk\cdot\rr_{ij}}-\\ \label{eq:ewald}
&&\frac{1}{\Omega}\sum_{i<j}\tilde{v}_{s0}q_iq_j
+\sum_iq_i^2v_M.
\eq

\begin{acknowledgments}
We would like to thank Saverio Moroni for several relevant discussions
at S.I.S.S.A. of Trieste and David Ceperley for many e-mail exchanges
which has been determinant for the realization of the new algorithm. 
\end{acknowledgments} 
\bibliography{jft}

\begin{thebibliography}{70}%
\makeatletter
\providecommand \@ifxundefined [1]{%
 \@ifx{#1\undefined}
}%
\providecommand \@ifnum [1]{%
 \ifnum #1\expandafter \@firstoftwo
 \else \expandafter \@secondoftwo
 \fi
}%
\providecommand \@ifx [1]{%
 \ifx #1\expandafter \@firstoftwo
 \else \expandafter \@secondoftwo
 \fi
}%
\providecommand \natexlab [1]{#1}%
\providecommand \enquote  [1]{``#1''}%
\providecommand \bibnamefont  [1]{#1}%
\providecommand \bibfnamefont [1]{#1}%
\providecommand \citenamefont [1]{#1}%
\providecommand \href@noop [0]{\@secondoftwo}%
\providecommand \href [0]{\begingroup \@sanitize@url \@href}%
\providecommand \@href[1]{\@@startlink{#1}\@@href}%
\providecommand \@@href[1]{\endgroup#1\@@endlink}%
\providecommand \@sanitize@url [0]{\catcode `\\12\catcode `\$12\catcode
  `\&12\catcode `\#12\catcode `\^12\catcode `\_12\catcode `\%12\relax}%
\providecommand \@@startlink[1]{}%
\providecommand \@@endlink[0]{}%
\providecommand \url  [0]{\begingroup\@sanitize@url \@url }%
\providecommand \@url [1]{\endgroup\@href {#1}{\urlprefix }}%
\providecommand \urlprefix  [0]{URL }%
\providecommand \Eprint [0]{\href }%
\providecommand \doibase [0]{https://doi.org/}%
\providecommand \selectlanguage [0]{\@gobble}%
\providecommand \bibinfo  [0]{\@secondoftwo}%
\providecommand \bibfield  [0]{\@secondoftwo}%
\providecommand \translation [1]{[#1]}%
\providecommand \BibitemOpen [0]{}%
\providecommand \bibitemStop [0]{}%
\providecommand \bibitemNoStop [0]{.\EOS\space}%
\providecommand \EOS [0]{\spacefactor3000\relax}%
\providecommand \BibitemShut  [1]{\csname bibitem#1\endcsname}%
\let\auto@bib@innerbib\@empty
\bibitem [{\citenamefont {Fantoni}(2013)}]{Fantoni2013}%
  \BibitemOpen
  \bibfield  {author} {\bibinfo {author} {\bibfnamefont {R.}~\bibnamefont
  {Fantoni}},\ }\href@noop {} {\bibfield  {journal} {\bibinfo  {journal} {Eur.
  Phys. J. B}\ }\textbf {\bibinfo {volume} {86}},\ \bibinfo {pages} {286}
  (\bibinfo {year} {2013})}\BibitemShut {NoStop}%
\bibitem [{\citenamefont {Fantoni}(2021)}]{Fantoni2021}%
  \BibitemOpen
  \bibfield  {author} {\bibinfo {author} {\bibfnamefont {R.}~\bibnamefont
  {Fantoni}},\ }\href@noop {} {\bibfield  {journal} {\bibinfo  {journal} {Eur.
  Phys. J. B}\ }\textbf {\bibinfo {volume} {94}},\ \bibinfo {pages} {63}
  (\bibinfo {year} {2021})}\BibitemShut {NoStop}%
\bibitem [{\citenamefont {Ashcroft}\ and\ \citenamefont
  {Mermin}(1976)}]{Ashcroft-Mermin}%
  \BibitemOpen
  \bibfield  {author} {\bibinfo {author} {\bibfnamefont {N.~W.}\ \bibnamefont
  {Ashcroft}}\ and\ \bibinfo {author} {\bibfnamefont {N.~D.}\ \bibnamefont
  {Mermin}},\ }\href@noop {} {\emph {\bibinfo {title} {Solid State Physics}}}\
  (\bibinfo  {publisher} {Harcourt, Inc.},\ \bibinfo {address} {Forth Worth},\
  \bibinfo {year} {1976})\BibitemShut {NoStop}%
\bibitem [{\citenamefont {Hansen}\ and\ \citenamefont
  {McDonald}(1986)}]{Hansen}%
  \BibitemOpen
  \bibfield  {author} {\bibinfo {author} {\bibfnamefont {J.~P.}\ \bibnamefont
  {Hansen}}\ and\ \bibinfo {author} {\bibfnamefont {I.~R.}\ \bibnamefont
  {McDonald}},\ }\href@noop {} {\emph {\bibinfo {title} {Theory of simple
  liquids}}},\ \bibinfo {edition} {2nd}\ ed.\ (\bibinfo  {publisher} {Academic
  Press},\ \bibinfo {address} {London},\ \bibinfo {year} {1986})\BibitemShut
  {NoStop}%
\bibitem [{\citenamefont {Shapiro}\ and\ \citenamefont
  {Teukolsky}(1983)}]{Shapiro-Teukolsky}%
  \BibitemOpen
  \bibfield  {author} {\bibinfo {author} {\bibfnamefont {S.~L.}\ \bibnamefont
  {Shapiro}}\ and\ \bibinfo {author} {\bibfnamefont {S.~A.}\ \bibnamefont
  {Teukolsky}},\ }\href@noop {} {\emph {\bibinfo {title} {Black Holes, White
  Dwarfs, and Neutron Stars. The Physics of Compact Objects}}}\ (\bibinfo
  {publisher} {John Wiley \& Sons, Inc.},\ \bibinfo {address} {Germany},\
  \bibinfo {year} {1983})\BibitemShut {NoStop}%
\bibitem [{\citenamefont {Fantoni}\ \emph {et~al.}(2003)\citenamefont
  {Fantoni}, \citenamefont {Jancovici},\ and\ \citenamefont
  {T\'{e}llez}}]{Fantoni03jsp}%
  \BibitemOpen
  \bibfield  {author} {\bibinfo {author} {\bibfnamefont {R.}~\bibnamefont
  {Fantoni}}, \bibinfo {author} {\bibfnamefont {B.}~\bibnamefont {Jancovici}},\
  and\ \bibinfo {author} {\bibfnamefont {G.}~\bibnamefont {T\'{e}llez}},\
  }\href@noop {} {\bibfield  {journal} {\bibinfo  {journal} {J. Stat. Phys.}\
  }\textbf {\bibinfo {volume} {{\bf 112}}},\ \bibinfo {pages} {27} (\bibinfo
  {year} {2003})}\BibitemShut {NoStop}%
\bibitem [{\citenamefont {Fantoni}\ and\ \citenamefont
  {T\'ellez}(2008)}]{Fantoni2008}%
  \BibitemOpen
  \bibfield  {author} {\bibinfo {author} {\bibfnamefont {R.}~\bibnamefont
  {Fantoni}}\ and\ \bibinfo {author} {\bibfnamefont {G.}~\bibnamefont
  {T\'ellez}},\ }\href@noop {} {\bibfield  {journal} {\bibinfo  {journal} {J.
  Stat. Phys.}\ }\textbf {\bibinfo {volume} {133}},\ \bibinfo {pages} {449}
  (\bibinfo {year} {2008})}\BibitemShut {NoStop}%
\bibitem [{\citenamefont {Fantoni}(2012{\natexlab{a}})}]{Fantoni2012}%
  \BibitemOpen
  \bibfield  {author} {\bibinfo {author} {\bibfnamefont {R.}~\bibnamefont
  {Fantoni}},\ }\href@noop {} {\bibfield  {journal} {\bibinfo  {journal} {J.
  Stat. Mech.}\ ,\ \bibinfo {pages} {P04015}} (\bibinfo {year}
  {2012}{\natexlab{a}})}\BibitemShut {NoStop}%
\bibitem [{\citenamefont {Fantoni}(2012{\natexlab{b}})}]{Fantoni2012b}%
  \BibitemOpen
  \bibfield  {author} {\bibinfo {author} {\bibfnamefont {R.}~\bibnamefont
  {Fantoni}},\ }\href@noop {} {\bibfield  {journal} {\bibinfo  {journal} {J.
  Stat. Mech.}\ ,\ \bibinfo {pages} {P10024}} (\bibinfo {year}
  {2012}{\natexlab{b}})}\BibitemShut {NoStop}%
\bibitem [{\citenamefont {Brown}\ \emph {et~al.}(2013)\citenamefont {Brown},
  \citenamefont {Clark}, \citenamefont {{DuBois}},\ and\ \citenamefont
  {Ceperley}}]{Brown2013}%
  \BibitemOpen
  \bibfield  {author} {\bibinfo {author} {\bibfnamefont {E.~W.}\ \bibnamefont
  {Brown}}, \bibinfo {author} {\bibfnamefont {B.~K.}\ \bibnamefont {Clark}},
  \bibinfo {author} {\bibfnamefont {J.~L.}\ \bibnamefont {{DuBois}}},\ and\
  \bibinfo {author} {\bibfnamefont {D.~M.}\ \bibnamefont {Ceperley}},\
  }\href@noop {} {\bibfield  {journal} {\bibinfo  {journal} {Phys. Rev. Lett.}\
  }\textbf {\bibinfo {volume} {110}},\ \bibinfo {pages} {146405} (\bibinfo
  {year} {2013})}\BibitemShut {NoStop}%
\bibitem [{\citenamefont {Brown}\ \emph {et~al.}(2014)\citenamefont {Brown},
  \citenamefont {Morales}, \citenamefont {Pierleoni},\ and\ \citenamefont
  {Ceperley}}]{Brown2014}%
  \BibitemOpen
  \bibfield  {author} {\bibinfo {author} {\bibfnamefont {E.}~\bibnamefont
  {Brown}}, \bibinfo {author} {\bibfnamefont {M.~A.}\ \bibnamefont {Morales}},
  \bibinfo {author} {\bibfnamefont {C.}~\bibnamefont {Pierleoni}},\ and\
  \bibinfo {author} {\bibfnamefont {D.~M.}\ \bibnamefont {Ceperley}},\
  }\bibfield  {title} {\bibinfo {title} {Quantum monte carlo techniques and
  applications for warm dense matter},\ }in\ \href@noop {} {\emph {\bibinfo
  {booktitle} {Frontiers and Challenges in Warm Dense Matter}}},\ \bibinfo
  {editor} {edited by\ \bibinfo {editor} {\bibnamefont {{F. Graziani {\sl et
  al.}}}}}\ (\bibinfo  {publisher} {Springer},\ \bibinfo {year} {2014})\ pp.\
  \bibinfo {pages} {123--149}\BibitemShut {NoStop}%
\bibitem [{\citenamefont {Schoof}\ \emph {et~al.}(2011)\citenamefont {Schoof},
  \citenamefont {Bonitz}, \citenamefont {Filinov}, \citenamefont {Hochsthul},\
  and\ \citenamefont {Dufty}}]{Schoof2011}%
  \BibitemOpen
  \bibfield  {author} {\bibinfo {author} {\bibfnamefont {T.}~\bibnamefont
  {Schoof}}, \bibinfo {author} {\bibfnamefont {M.}~\bibnamefont {Bonitz}},
  \bibinfo {author} {\bibfnamefont {A.}~\bibnamefont {Filinov}}, \bibinfo
  {author} {\bibfnamefont {D.}~\bibnamefont {Hochsthul}},\ and\ \bibinfo
  {author} {\bibfnamefont {J.~W.}\ \bibnamefont {Dufty}},\ }\bibfield  {title}
  {\bibinfo {title} {Configuration path integral monte carlo},\ }\href@noop {}
  {\bibfield  {journal} {\bibinfo  {journal} {Contrib. Plasma. Phys.}\ }\textbf
  {\bibinfo {volume} {51}},\ \bibinfo {pages} {687} (\bibinfo {year}
  {2011})}\BibitemShut {NoStop}%
\bibitem [{\citenamefont {Schoof}\ \emph {et~al.}(2015)\citenamefont {Schoof},
  \citenamefont {Groth}, \citenamefont {Vorberger},\ and\ \citenamefont
  {Bonitz}}]{Schoof2015}%
  \BibitemOpen
  \bibfield  {author} {\bibinfo {author} {\bibfnamefont {T.}~\bibnamefont
  {Schoof}}, \bibinfo {author} {\bibfnamefont {S.}~\bibnamefont {Groth}},
  \bibinfo {author} {\bibfnamefont {J.}~\bibnamefont {Vorberger}},\ and\
  \bibinfo {author} {\bibfnamefont {M.}~\bibnamefont {Bonitz}},\ }\href@noop {}
  {\bibfield  {journal} {\bibinfo  {journal} {Phys. Rev. Lett.}\ }\textbf
  {\bibinfo {volume} {115}},\ \bibinfo {pages} {130402} (\bibinfo {year}
  {2015})}\BibitemShut {NoStop}%
\bibitem [{\citenamefont {Dornheim}\ \emph {et~al.}(2015)\citenamefont
  {Dornheim}, \citenamefont {Groth}, \citenamefont {Filinov},\ and\
  \citenamefont {Bonitz}}]{Dornheim2015}%
  \BibitemOpen
  \bibfield  {author} {\bibinfo {author} {\bibfnamefont {T.}~\bibnamefont
  {Dornheim}}, \bibinfo {author} {\bibfnamefont {S.}~\bibnamefont {Groth}},
  \bibinfo {author} {\bibfnamefont {A.}~\bibnamefont {Filinov}},\ and\ \bibinfo
  {author} {\bibfnamefont {M.}~\bibnamefont {Bonitz}},\ }\bibfield  {title}
  {\bibinfo {title} {Permutation blocking path integral monte carlo: a highly
  efficient approach to the simulation of strongly degenerate non-ideal
  fermions},\ }\href@noop {} {\bibfield  {journal} {\bibinfo  {journal} {New J.
  Phys.}\ }\textbf {\bibinfo {volume} {17}},\ \bibinfo {pages} {073017}
  (\bibinfo {year} {2015})}\BibitemShut {NoStop}%
\bibitem [{\citenamefont {Dornheim}\ \emph
  {et~al.}(2016{\natexlab{a}})\citenamefont {Dornheim}, \citenamefont {Groth},
  \citenamefont {Sjostrom}, \citenamefont {Malone}, \citenamefont {Foulkes},\
  and\ \citenamefont {Bonitz}}]{Dornheim2016}%
  \BibitemOpen
  \bibfield  {author} {\bibinfo {author} {\bibfnamefont {T.}~\bibnamefont
  {Dornheim}}, \bibinfo {author} {\bibfnamefont {S.}~\bibnamefont {Groth}},
  \bibinfo {author} {\bibfnamefont {T.}~\bibnamefont {Sjostrom}}, \bibinfo
  {author} {\bibfnamefont {F.~D.}\ \bibnamefont {Malone}}, \bibinfo {author}
  {\bibfnamefont {W.~M.~C.}\ \bibnamefont {Foulkes}},\ and\ \bibinfo {author}
  {\bibfnamefont {M.}~\bibnamefont {Bonitz}},\ }\href@noop {} {\bibfield
  {journal} {\bibinfo  {journal} {Phys. Rev. Lett.}\ }\textbf {\bibinfo
  {volume} {117}},\ \bibinfo {pages} {156403} (\bibinfo {year}
  {2016}{\natexlab{a}})}\BibitemShut {NoStop}%
\bibitem [{\citenamefont {Groth}\ \emph
  {et~al.}(2017{\natexlab{a}})\citenamefont {Groth}, \citenamefont {Dornheim},
  \citenamefont {Sjostrom}, \citenamefont {Malone}, \citenamefont {Foulkes},\
  and\ \citenamefont {Bonitz}}]{Groth2017}%
  \BibitemOpen
  \bibfield  {author} {\bibinfo {author} {\bibfnamefont {S.}~\bibnamefont
  {Groth}}, \bibinfo {author} {\bibfnamefont {T.}~\bibnamefont {Dornheim}},
  \bibinfo {author} {\bibfnamefont {T.}~\bibnamefont {Sjostrom}}, \bibinfo
  {author} {\bibfnamefont {F.~D.}\ \bibnamefont {Malone}}, \bibinfo {author}
  {\bibfnamefont {W.~M.~C.}\ \bibnamefont {Foulkes}},\ and\ \bibinfo {author}
  {\bibfnamefont {M.}~\bibnamefont {Bonitz}},\ }\href@noop {} {\bibfield
  {journal} {\bibinfo  {journal} {Phys. Rev. Lett.}\ }\textbf {\bibinfo
  {volume} {119}},\ \bibinfo {pages} {135001} (\bibinfo {year}
  {2017}{\natexlab{a}})}\BibitemShut {NoStop}%
\bibitem [{\citenamefont {Malone}\ \emph {et~al.}(2016)\citenamefont {Malone},
  \citenamefont {Blunt}, \citenamefont {Brown}, \citenamefont {Lee},
  \citenamefont {Spencer}, \citenamefont {Foulkes},\ and\ \citenamefont
  {Shepherd}}]{Malone2016}%
  \BibitemOpen
  \bibfield  {author} {\bibinfo {author} {\bibfnamefont {F.~D.}\ \bibnamefont
  {Malone}}, \bibinfo {author} {\bibfnamefont {N.~S.}\ \bibnamefont {Blunt}},
  \bibinfo {author} {\bibfnamefont {E.~W.}\ \bibnamefont {Brown}}, \bibinfo
  {author} {\bibfnamefont {D.~K.~K.}\ \bibnamefont {Lee}}, \bibinfo {author}
  {\bibfnamefont {J.~S.}\ \bibnamefont {Spencer}}, \bibinfo {author}
  {\bibfnamefont {W.~M.~C.}\ \bibnamefont {Foulkes}},\ and\ \bibinfo {author}
  {\bibfnamefont {J.~J.}\ \bibnamefont {Shepherd}},\ }\bibfield  {title}
  {\bibinfo {title} {Accurate exchange-correlation energies for the warm dense
  electron gas},\ }\href@noop {} {\bibfield  {journal} {\bibinfo  {journal}
  {Phys. Rev. Lett.}\ }\textbf {\bibinfo {volume} {117}},\ \bibinfo {pages}
  {115701} (\bibinfo {year} {2016})}\BibitemShut {NoStop}%
\bibitem [{\citenamefont {Filinov}\ \emph {et~al.}(2015)\citenamefont
  {Filinov}, \citenamefont {Fortov}, \citenamefont {Bonitz},\ and\
  \citenamefont {Moldabekov}}]{Filinov2015}%
  \BibitemOpen
  \bibfield  {author} {\bibinfo {author} {\bibfnamefont {V.~S.}\ \bibnamefont
  {Filinov}}, \bibinfo {author} {\bibfnamefont {V.~E.}\ \bibnamefont {Fortov}},
  \bibinfo {author} {\bibfnamefont {M.}~\bibnamefont {Bonitz}},\ and\ \bibinfo
  {author} {\bibfnamefont {Z.}~\bibnamefont {Moldabekov}},\ }\bibfield  {title}
  {\bibinfo {title} {Fermionic path-integral monte carlo results for the
  uniform electron gas at finite temperature},\ }\href@noop {} {\bibfield
  {journal} {\bibinfo  {journal} {Phys. Rev. E}\ }\textbf {\bibinfo {volume}
  {91}},\ \bibinfo {pages} {033108} (\bibinfo {year} {2015})}\BibitemShut
  {NoStop}%
\bibitem [{\citenamefont {Fantoni}(2018{\natexlab{a}})}]{Fantoni2018b}%
  \BibitemOpen
  \bibfield  {author} {\bibinfo {author} {\bibfnamefont {R.}~\bibnamefont
  {Fantoni}},\ }\href@noop {} {\bibfield  {journal} {\bibinfo  {journal} {Int.
  J. Mod. Phys. C}\ }\textbf {\bibinfo {volume} {29}},\ \bibinfo {pages}
  {1850028} (\bibinfo {year} {2018}{\natexlab{a}})}\BibitemShut {NoStop}%
\bibitem [{\citenamefont {March}\ and\ \citenamefont
  {Tosi}(1984)}]{March-Tosi}%
  \BibitemOpen
  \bibfield  {author} {\bibinfo {author} {\bibfnamefont {N.~H.}\ \bibnamefont
  {March}}\ and\ \bibinfo {author} {\bibfnamefont {M.~P.}\ \bibnamefont
  {Tosi}},\ }\href@noop {} {\emph {\bibinfo {title} {Coulomb Liquids}}}\
  (\bibinfo  {publisher} {Academic Press},\ \bibinfo {address} {London},\
  \bibinfo {year} {1984})\BibitemShut {NoStop}%
\bibitem [{\citenamefont {Friedel}(1958)}]{Friedel1958}%
  \BibitemOpen
  \bibfield  {author} {\bibinfo {author} {\bibfnamefont {J.}~\bibnamefont
  {Friedel}},\ }\href@noop {} {\bibfield  {journal} {\bibinfo  {journal} {N.
  Cimento Suppl.}\ }\textbf {\bibinfo {volume} {7}},\ \bibinfo {pages} {287}
  (\bibinfo {year} {1958})}\BibitemShut {NoStop}%
\bibitem [{\citenamefont {Lighthill}(1959)}]{Lighthill}%
  \BibitemOpen
  \bibfield  {author} {\bibinfo {author} {\bibfnamefont {M.~J.}\ \bibnamefont
  {Lighthill}},\ }\href@noop {} {\emph {\bibinfo {title} {Introduction to
  Fourier Analysis and Generalized Functions}}}\ (\bibinfo  {publisher}
  {Cambridge University Press},\ \bibinfo {year} {1959})\ \bibinfo {note}
  {theorem 19}\BibitemShut {NoStop}%
\bibitem [{\citenamefont {{D. M. Ceperley}}(1995)}]{Ceperley1995}%
  \BibitemOpen
  \bibfield  {author} {\bibinfo {author} {\bibnamefont {{D. M. Ceperley}}},\
  }\href@noop {} {\bibfield  {journal} {\bibinfo  {journal} {Rev. Mod. Phys.}\
  }\textbf {\bibinfo {volume} {67}},\ \bibinfo {pages} {279} (\bibinfo {year}
  {1995})}\BibitemShut {NoStop}%
\bibitem [{\citenamefont {Hansen}(1973)}]{Hansen1973}%
  \BibitemOpen
  \bibfield  {author} {\bibinfo {author} {\bibfnamefont {J.~P.}\ \bibnamefont
  {Hansen}},\ }\href@noop {} {\bibfield  {journal} {\bibinfo  {journal} {Phys.
  Rev. A}\ }\textbf {\bibinfo {volume} {8}},\ \bibinfo {pages} {3096} (\bibinfo
  {year} {1973})}\BibitemShut {NoStop}%
\bibitem [{\citenamefont {Hansen}\ and\ \citenamefont
  {Vieillefosse}(1975)}]{Hansen1975}%
  \BibitemOpen
  \bibfield  {author} {\bibinfo {author} {\bibfnamefont {J.~P.}\ \bibnamefont
  {Hansen}}\ and\ \bibinfo {author} {\bibfnamefont {P.}~\bibnamefont
  {Vieillefosse}},\ }\href@noop {} {\bibfield  {journal} {\bibinfo  {journal}
  {Phys. Lett.}\ }\textbf {\bibinfo {volume} {53A}},\ \bibinfo {pages} {187}
  (\bibinfo {year} {1975})}\BibitemShut {NoStop}%
\bibitem [{\citenamefont {Gupta}\ and\ \citenamefont
  {Rajagopal}(1980)}]{Gupta1980}%
  \BibitemOpen
  \bibfield  {author} {\bibinfo {author} {\bibfnamefont {U.}~\bibnamefont
  {Gupta}}\ and\ \bibinfo {author} {\bibfnamefont {A.~K.}\ \bibnamefont
  {Rajagopal}},\ }\href@noop {} {\bibfield  {journal} {\bibinfo  {journal}
  {Phys. Rev. A}\ }\textbf {\bibinfo {volume} {22}},\ \bibinfo {pages} {2792}
  (\bibinfo {year} {1980})}\BibitemShut {NoStop}%
\bibitem [{\citenamefont {Perrot}\ and\ \citenamefont
  {Dharma-wardana}(1984)}]{Perrot1984}%
  \BibitemOpen
  \bibfield  {author} {\bibinfo {author} {\bibfnamefont {F.}~\bibnamefont
  {Perrot}}\ and\ \bibinfo {author} {\bibfnamefont {M.~W.~C.}\ \bibnamefont
  {Dharma-wardana}},\ }\href@noop {} {\bibfield  {journal} {\bibinfo  {journal}
  {Phys. Rev. A}\ }\textbf {\bibinfo {volume} {30}},\ \bibinfo {pages} {2619}
  (\bibinfo {year} {1984})}\BibitemShut {NoStop}%
\bibitem [{\citenamefont {Singwi}\ \emph {et~al.}(1968)\citenamefont {Singwi},
  \citenamefont {Tosi}, \citenamefont {Land},\ and\ \citenamefont
  {Sj\"olander}}]{Singwi1968}%
  \BibitemOpen
  \bibfield  {author} {\bibinfo {author} {\bibfnamefont {K.~S.}\ \bibnamefont
  {Singwi}}, \bibinfo {author} {\bibfnamefont {M.~P.}\ \bibnamefont {Tosi}},
  \bibinfo {author} {\bibfnamefont {R.~H.}\ \bibnamefont {Land}},\ and\
  \bibinfo {author} {\bibfnamefont {A.}~\bibnamefont {Sj\"olander}},\
  }\href@noop {} {\bibfield  {journal} {\bibinfo  {journal} {Phys. Rev.}\
  }\textbf {\bibinfo {volume} {176}},\ \bibinfo {pages} {589} (\bibinfo {year}
  {1968})}\BibitemShut {NoStop}%
\bibitem [{\citenamefont {Tanaka}\ and\ \citenamefont
  {Ichimaru}(1986)}]{Tanaka1986}%
  \BibitemOpen
  \bibfield  {author} {\bibinfo {author} {\bibfnamefont {S.}~\bibnamefont
  {Tanaka}}\ and\ \bibinfo {author} {\bibfnamefont {S.}~\bibnamefont
  {Ichimaru}},\ }\href@noop {} {\bibfield  {journal} {\bibinfo  {journal}
  {Journal of the Physical Society of Japan}\ }\textbf {\bibinfo {volume}
  {55}},\ \bibinfo {pages} {2278} (\bibinfo {year} {1986})}\BibitemShut
  {NoStop}%
\bibitem [{\citenamefont {Perrot}\ and\ \citenamefont
  {Dharma-wardana}(2000)}]{Perrot2000}%
  \BibitemOpen
  \bibfield  {author} {\bibinfo {author} {\bibfnamefont {F.~M.~C.}\
  \bibnamefont {Perrot}}\ and\ \bibinfo {author} {\bibfnamefont {M.~W.~C.}\
  \bibnamefont {Dharma-wardana}},\ }\href@noop {} {\bibfield  {journal}
  {\bibinfo  {journal} {Phys. Rev. B}\ }\textbf {\bibinfo {volume} {62}},\
  \bibinfo {pages} {16536} (\bibinfo {year} {2000})}\BibitemShut {NoStop}%
\bibitem [{\citenamefont {Dharma-wardana}\ and\ \citenamefont
  {Perrot}(2000)}]{Dharma2000}%
  \BibitemOpen
  \bibfield  {author} {\bibinfo {author} {\bibfnamefont {M.~W.~C.}\
  \bibnamefont {Dharma-wardana}}\ and\ \bibinfo {author} {\bibfnamefont
  {F.}~\bibnamefont {Perrot}},\ }\href@noop {} {\bibfield  {journal} {\bibinfo
  {journal} {Phys. Rev. Lett.}\ }\textbf {\bibinfo {volume} {84}},\ \bibinfo
  {pages} {959} (\bibinfo {year} {2000})}\BibitemShut {NoStop}%
\bibitem [{\citenamefont {Dutta}\ and\ \citenamefont
  {Dufty}(2013)}]{Dutta2012}%
  \BibitemOpen
  \bibfield  {author} {\bibinfo {author} {\bibfnamefont {S.}~\bibnamefont
  {Dutta}}\ and\ \bibinfo {author} {\bibfnamefont {J.}~\bibnamefont {Dufty}},\
  }\href@noop {} {\bibfield  {journal} {\bibinfo  {journal} {Phys. Rev. E}\
  }\textbf {\bibinfo {volume} {87}},\ \bibinfo {pages} {032102} (\bibinfo
  {year} {2013})}\BibitemShut {NoStop}%
\bibitem [{\citenamefont {Groth}\ \emph
  {et~al.}(2017{\natexlab{b}})\citenamefont {Groth}, \citenamefont {Dornheim},\
  and\ \citenamefont {Bonitz}}]{Groth2017b}%
  \BibitemOpen
  \bibfield  {author} {\bibinfo {author} {\bibfnamefont {S.}~\bibnamefont
  {Groth}}, \bibinfo {author} {\bibfnamefont {T.}~\bibnamefont {Dornheim}},\
  and\ \bibinfo {author} {\bibfnamefont {M.}~\bibnamefont {Bonitz}},\
  }\bibfield  {title} {\bibinfo {title} {Free energy of the uniform electron
  gas: Testing analytical models against first-principles results},\
  }\href@noop {} {\bibfield  {journal} {\bibinfo  {journal} {Contrib. Plasma
  Phys.}\ }\textbf {\bibinfo {volume} {57}},\ \bibinfo {pages} {137} (\bibinfo
  {year} {2017}{\natexlab{b}})}\BibitemShut {NoStop}%
\bibitem [{\citenamefont {Karasiev}\ \emph {et~al.}(2014)\citenamefont
  {Karasiev}, \citenamefont {Sjostrom}, \citenamefont {Dufty},\ and\
  \citenamefont {Trickey}}]{Karasiev2014}%
  \BibitemOpen
  \bibfield  {author} {\bibinfo {author} {\bibfnamefont {V.~V.}\ \bibnamefont
  {Karasiev}}, \bibinfo {author} {\bibfnamefont {T.}~\bibnamefont {Sjostrom}},
  \bibinfo {author} {\bibfnamefont {J.}~\bibnamefont {Dufty}},\ and\ \bibinfo
  {author} {\bibfnamefont {S.}~\bibnamefont {Trickey}},\ }\href@noop {}
  {\bibfield  {journal} {\bibinfo  {journal} {Phys. Rev. Lett.}\ }\textbf
  {\bibinfo {volume} {112}},\ \bibinfo {pages} {076403} (\bibinfo {year}
  {2014})}\BibitemShut {NoStop}%
\bibitem [{\citenamefont {Karasiev}\ \emph {et~al.}(2019)\citenamefont
  {Karasiev}, \citenamefont {Trickey},\ and\ \citenamefont
  {Dufty}}]{Karasiev2019}%
  \BibitemOpen
  \bibfield  {author} {\bibinfo {author} {\bibfnamefont {V.~V.}\ \bibnamefont
  {Karasiev}}, \bibinfo {author} {\bibfnamefont {S.~B.}\ \bibnamefont
  {Trickey}},\ and\ \bibinfo {author} {\bibfnamefont {J.~W.}\ \bibnamefont
  {Dufty}},\ }\href@noop {} {\bibfield  {journal} {\bibinfo  {journal} {Phys.
  Rev. B}\ }\textbf {\bibinfo {volume} {99}},\ \bibinfo {pages} {195134}
  (\bibinfo {year} {2019})}\BibitemShut {NoStop}%
\bibitem [{\citenamefont {Ceperley}(1991)}]{Ceperley1991}%
  \BibitemOpen
  \bibfield  {author} {\bibinfo {author} {\bibfnamefont {D.~M.}\ \bibnamefont
  {Ceperley}},\ }\href@noop {} {\bibfield  {journal} {\bibinfo  {journal} {J.
  Stat. Phys.}\ }\textbf {\bibinfo {volume} {63}},\ \bibinfo {pages} {1237}
  (\bibinfo {year} {1991})}\BibitemShut {NoStop}%
\bibitem [{\citenamefont {Ceperley}(1996)}]{Ceperley1996}%
  \BibitemOpen
  \bibfield  {author} {\bibinfo {author} {\bibfnamefont {D.~M.}\ \bibnamefont
  {Ceperley}},\ }\bibfield  {title} {\bibinfo {title} {Path integral monte
  carlo methods for fermions},\ }in\ \href@noop {} {\emph {\bibinfo {booktitle}
  {Monte Carlo and Molecular Dynamics of Condensed Matter Systems}}},\ \bibinfo
  {editor} {edited by\ \bibinfo {editor} {\bibfnamefont {K.}~\bibnamefont
  {Binder}}\ and\ \bibinfo {editor} {\bibfnamefont {G.}~\bibnamefont
  {Ciccotti}}}\ (\bibinfo  {publisher} {Editrice Compositori},\ \bibinfo
  {address} {Bologna, Italy},\ \bibinfo {year} {1996})\BibitemShut {NoStop}%
\bibitem [{\citenamefont {Prokof'ev}\ \emph {et~al.}(1998)\citenamefont
  {Prokof'ev}, \citenamefont {Svistunov},\ and\ \citenamefont
  {Tupitsyn}}]{Prokofev1998}%
  \BibitemOpen
  \bibfield  {author} {\bibinfo {author} {\bibfnamefont {N.~V.}\ \bibnamefont
  {Prokof'ev}}, \bibinfo {author} {\bibfnamefont {B.~V.}\ \bibnamefont
  {Svistunov}},\ and\ \bibinfo {author} {\bibfnamefont {I.~S.}\ \bibnamefont
  {Tupitsyn}},\ }\bibfield  {title} {\bibinfo {title} {Exact, complete, and
  universal continuoustime worldline monte carlo approach to the statistics of
  discrete quantum systems},\ }\href@noop {} {\bibfield  {journal} {\bibinfo
  {journal} {J. Exp. Theor. Phys.}\ }\textbf {\bibinfo {volume} {87}},\
  \bibinfo {pages} {310} (\bibinfo {year} {1998})}\BibitemShut {NoStop}%
\bibitem [{\citenamefont {Boninsegni}\ \emph
  {et~al.}(2006{\natexlab{a}})\citenamefont {Boninsegni}, \citenamefont
  {Prokof'ev},\ and\ \citenamefont {Svistunov}}]{Boninsegni2006a}%
  \BibitemOpen
  \bibfield  {author} {\bibinfo {author} {\bibfnamefont {M.}~\bibnamefont
  {Boninsegni}}, \bibinfo {author} {\bibfnamefont {N.}~\bibnamefont
  {Prokof'ev}},\ and\ \bibinfo {author} {\bibfnamefont {B.}~\bibnamefont
  {Svistunov}},\ }\href@noop {} {\bibfield  {journal} {\bibinfo  {journal}
  {Phys. Rev. Lett.}\ }\textbf {\bibinfo {volume} {96}},\ \bibinfo {pages}
  {070601} (\bibinfo {year} {2006}{\natexlab{a}})}\BibitemShut {NoStop}%
\bibitem [{\citenamefont {Singwi}\ and\ \citenamefont
  {Tosi}(1981)}]{Singwi1981}%
  \BibitemOpen
  \bibfield  {author} {\bibinfo {author} {\bibfnamefont {K.~S.}\ \bibnamefont
  {Singwi}}\ and\ \bibinfo {author} {\bibfnamefont {M.~P.}\ \bibnamefont
  {Tosi}},\ }\href@noop {} {\bibfield  {journal} {\bibinfo  {journal} {Sol.
  State Phys.}\ }\textbf {\bibinfo {volume} {36}},\ \bibinfo {pages} {177}
  (\bibinfo {year} {1981})}\BibitemShut {NoStop}%
\bibitem [{\citenamefont {Ichimaru}(1982)}]{Ichimaru1982}%
  \BibitemOpen
  \bibfield  {author} {\bibinfo {author} {\bibfnamefont {S.}~\bibnamefont
  {Ichimaru}},\ }\href@noop {} {\bibfield  {journal} {\bibinfo  {journal} {Rev.
  Mod. Phys.}\ }\textbf {\bibinfo {volume} {54}},\ \bibinfo {pages} {1017}
  (\bibinfo {year} {1982})}\BibitemShut {NoStop}%
\bibitem [{\citenamefont {Martin}(1988)}]{Martin88}%
  \BibitemOpen
  \bibfield  {author} {\bibinfo {author} {\bibfnamefont {P.~A.}\ \bibnamefont
  {Martin}},\ }\href@noop {} {\bibfield  {journal} {\bibinfo  {journal} {Rev.
  Mod. Phys.}\ }\textbf {\bibinfo {volume} {60}},\ \bibinfo {pages} {1075}
  (\bibinfo {year} {1988})}\BibitemShut {NoStop}%
\bibitem [{\citenamefont {Wigner}(1934)}]{Wigner1934}%
  \BibitemOpen
  \bibfield  {author} {\bibinfo {author} {\bibfnamefont {E.}~\bibnamefont
  {Wigner}},\ }\href@noop {} {\bibfield  {journal} {\bibinfo  {journal} {Phys.
  Rev.}\ }\textbf {\bibinfo {volume} {46}},\ \bibinfo {pages} {1002} (\bibinfo
  {year} {1934})}\BibitemShut {NoStop}%
\bibitem [{\citenamefont {Leggett}(1975)}]{Leggett1975}%
  \BibitemOpen
  \bibfield  {author} {\bibinfo {author} {\bibfnamefont {A.~J.}\ \bibnamefont
  {Leggett}},\ }\href@noop {} {\bibfield  {journal} {\bibinfo  {journal} {Rev.
  Mod. Phys.}\ }\textbf {\bibinfo {volume} {47}},\ \bibinfo {pages} {331}
  (\bibinfo {year} {1975})}\BibitemShut {NoStop}%
\bibitem [{\citenamefont {Giuliani}\ and\ \citenamefont
  {Vignale}(2005)}]{Giuliani-Vignale}%
  \BibitemOpen
  \bibfield  {author} {\bibinfo {author} {\bibfnamefont {G.~F.}\ \bibnamefont
  {Giuliani}}\ and\ \bibinfo {author} {\bibfnamefont {G.}~\bibnamefont
  {Vignale}},\ }\href@noop {} {\emph {\bibinfo {title} {Quantum Theory of the
  Electron Liquid}}}\ (\bibinfo  {publisher} {Cambridge University Press},\
  \bibinfo {address} {Cambridge},\ \bibinfo {year} {2005})\BibitemShut
  {NoStop}%
\bibitem [{\citenamefont {Pollock}\ and\ \citenamefont
  {Ceperley}(1987)}]{Pollock1987}%
  \BibitemOpen
  \bibfield  {author} {\bibinfo {author} {\bibfnamefont {E.~L.}\ \bibnamefont
  {Pollock}}\ and\ \bibinfo {author} {\bibfnamefont {D.~M.}\ \bibnamefont
  {Ceperley}},\ }\href@noop {} {\bibfield  {journal} {\bibinfo  {journal}
  {Phys. Rev. B}\ }\textbf {\bibinfo {volume} {36}},\ \bibinfo {pages} {8343}
  (\bibinfo {year} {1987})}\BibitemShut {NoStop}%
\bibitem [{\citenamefont {Ceperley}\ and\ \citenamefont
  {Alder}(1980)}]{Ceperley1980}%
  \BibitemOpen
  \bibfield  {author} {\bibinfo {author} {\bibfnamefont {D.~M.}\ \bibnamefont
  {Ceperley}}\ and\ \bibinfo {author} {\bibfnamefont {B.~J.}\ \bibnamefont
  {Alder}},\ }\href@noop {} {\bibfield  {journal} {\bibinfo  {journal} {Phys.
  Rev. Lett.}\ }\textbf {\bibinfo {volume} {45}},\ \bibinfo {pages} {566}
  (\bibinfo {year} {1980})}\BibitemShut {NoStop}%
\bibitem [{\citenamefont {Feynman}(1953{\natexlab{a}})}]{Feynman1953a}%
  \BibitemOpen
  \bibfield  {author} {\bibinfo {author} {\bibfnamefont {R.~P.}\ \bibnamefont
  {Feynman}},\ }\href@noop {} {\bibfield  {journal} {\bibinfo  {journal} {Phys.
  Rev.}\ }\textbf {\bibinfo {volume} {90}},\ \bibinfo {pages} {1116} (\bibinfo
  {year} {1953}{\natexlab{a}})}\BibitemShut {NoStop}%
\bibitem [{\citenamefont {Feynman}(1953{\natexlab{b}})}]{Feynman1953b}%
  \BibitemOpen
  \bibfield  {author} {\bibinfo {author} {\bibfnamefont {R.~P.}\ \bibnamefont
  {Feynman}},\ }\href@noop {} {\bibfield  {journal} {\bibinfo  {journal} {Phys.
  Rev.}\ }\textbf {\bibinfo {volume} {91}},\ \bibinfo {pages} {1291} (\bibinfo
  {year} {1953}{\natexlab{b}})}\BibitemShut {NoStop}%
\bibitem [{\citenamefont {Feynman}(1953{\natexlab{c}})}]{Feynman1953c}%
  \BibitemOpen
  \bibfield  {author} {\bibinfo {author} {\bibfnamefont {R.~P.}\ \bibnamefont
  {Feynman}},\ }\href@noop {} {\bibfield  {journal} {\bibinfo  {journal} {Phys.
  Rev.}\ }\textbf {\bibinfo {volume} {90}},\ \bibinfo {pages} {1301} (\bibinfo
  {year} {1953}{\natexlab{c}})}\BibitemShut {NoStop}%
\bibitem [{\citenamefont {Feynman}\ and\ \citenamefont
  {Hibbs}(1965)}]{Feynman-Hibbs}%
  \BibitemOpen
  \bibfield  {author} {\bibinfo {author} {\bibfnamefont {R.~P.}\ \bibnamefont
  {Feynman}}\ and\ \bibinfo {author} {\bibfnamefont {A.~R.}\ \bibnamefont
  {Hibbs}},\ }\href@noop {} {\emph {\bibinfo {title} {Quantum Mechanics and
  Path Integrals}}}\ (\bibinfo  {publisher} {McGraw-Hill Publishing Company},\
  \bibinfo {address} {New York},\ \bibinfo {year} {1965})\ \bibinfo {note}
  {page 292-293}\BibitemShut {NoStop}%
\bibitem [{\citenamefont {Metropolis}\ \emph {et~al.}(1953)\citenamefont
  {Metropolis}, \citenamefont {Rosenbluth}, \citenamefont {Rosenbluth},
  \citenamefont {Teller},\ and\ \citenamefont {Teller}}]{Metropolis}%
  \BibitemOpen
  \bibfield  {author} {\bibinfo {author} {\bibfnamefont {N.}~\bibnamefont
  {Metropolis}}, \bibinfo {author} {\bibfnamefont {A.~W.}\ \bibnamefont
  {Rosenbluth}}, \bibinfo {author} {\bibfnamefont {M.~N.}\ \bibnamefont
  {Rosenbluth}}, \bibinfo {author} {\bibfnamefont {A.~M.}\ \bibnamefont
  {Teller}},\ and\ \bibinfo {author} {\bibfnamefont {E.}~\bibnamefont
  {Teller}},\ }\href@noop {} {\bibfield  {journal} {\bibinfo  {journal} {J.
  Chem. Phys.}\ }\textbf {\bibinfo {volume} {1087}},\ \bibinfo {pages} {21}
  (\bibinfo {year} {1953})}\BibitemShut {NoStop}%
\bibitem [{\citenamefont {Pollock}(1988)}]{Pollock1988}%
  \BibitemOpen
  \bibfield  {author} {\bibinfo {author} {\bibfnamefont {E.~L.}\ \bibnamefont
  {Pollock}},\ }\href@noop {} {\bibfield  {journal} {\bibinfo  {journal}
  {Computer Physics Communications}\ }\textbf {\bibinfo {volume} {52}},\
  \bibinfo {pages} {49} (\bibinfo {year} {1988})}\BibitemShut {NoStop}%
\bibitem [{\citenamefont {Vieillefosse}(1994)}]{Vieillefosse1994}%
  \BibitemOpen
  \bibfield  {author} {\bibinfo {author} {\bibfnamefont {P.}~\bibnamefont
  {Vieillefosse}},\ }\href@noop {} {\bibfield  {journal} {\bibinfo  {journal}
  {J. Stat. Phys.}\ }\textbf {\bibinfo {volume} {74}},\ \bibinfo {pages} {1195}
  (\bibinfo {year} {1994})}\BibitemShut {NoStop}%
\bibitem [{\citenamefont {Vieillefosse}(1995)}]{Vieillefosse1995}%
  \BibitemOpen
  \bibfield  {author} {\bibinfo {author} {\bibfnamefont {P.}~\bibnamefont
  {Vieillefosse}},\ }\href@noop {} {\bibfield  {journal} {\bibinfo  {journal}
  {J. Stat. Phys.}\ }\textbf {\bibinfo {volume} {80}},\ \bibinfo {pages} {461}
  (\bibinfo {year} {1995})}\BibitemShut {NoStop}%
\bibitem [{\citenamefont {Boninsegni}\ \emph
  {et~al.}(2006{\natexlab{b}})\citenamefont {Boninsegni}, \citenamefont
  {Prokof'ev},\ and\ \citenamefont {Svistunov}}]{Boninsegni2006b}%
  \BibitemOpen
  \bibfield  {author} {\bibinfo {author} {\bibfnamefont {M.}~\bibnamefont
  {Boninsegni}}, \bibinfo {author} {\bibfnamefont {N.~V.}\ \bibnamefont
  {Prokof'ev}},\ and\ \bibinfo {author} {\bibfnamefont {B.~V.}\ \bibnamefont
  {Svistunov}},\ }\href@noop {} {\bibfield  {journal} {\bibinfo  {journal}
  {Phys. Rev. E}\ }\textbf {\bibinfo {volume} {74}},\ \bibinfo {pages} {036701}
  (\bibinfo {year} {2006}{\natexlab{b}})}\BibitemShut {NoStop}%
\bibitem [{\citenamefont {Natoli}\ and\ \citenamefont
  {Ceperley}(1995)}]{Natoli1995}%
  \BibitemOpen
  \bibfield  {author} {\bibinfo {author} {\bibfnamefont {V.~D.}\ \bibnamefont
  {Natoli}}\ and\ \bibinfo {author} {\bibfnamefont {D.~M.}\ \bibnamefont
  {Ceperley}},\ }\bibfield  {title} {\bibinfo {title} {An optimized method for
  treating long-range potentials},\ }\href@noop {} {\bibfield  {journal}
  {\bibinfo  {journal} {J. Comput. Physics}\ }\textbf {\bibinfo {volume}
  {117}},\ \bibinfo {pages} {171} (\bibinfo {year} {1995})}\BibitemShut
  {NoStop}%
\bibitem [{\citenamefont {Fraser}\ \emph {et~al.}(1996)\citenamefont {Fraser},
  \citenamefont {Foulkes}, \citenamefont {Rajagopal}, \citenamefont {Needs},
  \citenamefont {Kenney},\ and\ \citenamefont {Williamson}}]{Fraser1996}%
  \BibitemOpen
  \bibfield  {author} {\bibinfo {author} {\bibfnamefont {L.~M.}\ \bibnamefont
  {Fraser}}, \bibinfo {author} {\bibfnamefont {W.~M.~C.}\ \bibnamefont
  {Foulkes}}, \bibinfo {author} {\bibfnamefont {G.}~\bibnamefont {Rajagopal}},
  \bibinfo {author} {\bibfnamefont {R.~J.}\ \bibnamefont {Needs}}, \bibinfo
  {author} {\bibfnamefont {S.~D.}\ \bibnamefont {Kenney}},\ and\ \bibinfo
  {author} {\bibfnamefont {A.~J.}\ \bibnamefont {Williamson}},\ }\bibfield
  {title} {\bibinfo {title} {Finite-size effects and coulomb interactions in
  quantum monte carlo calculations for homogeneous systems with periodic
  boundary conditions},\ }\href@noop {} {\bibfield  {journal} {\bibinfo
  {journal} {Phys. Rev. B}\ }\textbf {\bibinfo {volume} {53}},\ \bibinfo
  {pages} {1814} (\bibinfo {year} {1996})}\BibitemShut {NoStop}%
\bibitem [{\citenamefont {Fantoni}\ and\ \citenamefont
  {Moroni}(2014)}]{Fantoni2014b}%
  \BibitemOpen
  \bibfield  {author} {\bibinfo {author} {\bibfnamefont {R.}~\bibnamefont
  {Fantoni}}\ and\ \bibinfo {author} {\bibfnamefont {S.}~\bibnamefont
  {Moroni}},\ }\href@noop {} {\bibfield  {journal} {\bibinfo  {journal} {J.
  Chem. Phys.}\ }\textbf {\bibinfo {volume} {141}},\ \bibinfo {pages} {114110}
  (\bibinfo {year} {2014})}\BibitemShut {NoStop}%
\bibitem [{\citenamefont {Fantoni}(2015)}]{Fantoni15b}%
  \BibitemOpen
  \bibfield  {author} {\bibinfo {author} {\bibfnamefont {R.}~\bibnamefont
  {Fantoni}},\ }\bibfield  {title} {\bibinfo {title} {Two-phase coexistence for
  hydrogen-helium mixtures},\ }\href
  {https://doi.org/10.1103/PhysRevE.92.012133} {\bibfield  {journal} {\bibinfo
  {journal} {Phys. Rev. E}\ }\textbf {\bibinfo {volume} {92}},\ \bibinfo
  {pages} {012133} (\bibinfo {year} {2015})}\BibitemShut {NoStop}%
\bibitem [{\citenamefont {Fantoni}(2016)}]{Fantoni16a}%
  \BibitemOpen
  \bibfield  {author} {\bibinfo {author} {\bibfnamefont {R.}~\bibnamefont
  {Fantoni}},\ }\bibfield  {title} {\bibinfo {title} {Supercooled superfluids
  in monte carlo simulations},\ }\href
  {https://doi.org/10.1140/epjb/e2016-60917-9} {\bibfield  {journal} {\bibinfo
  {journal} {Eur. Phys. J. B}\ }\textbf {\bibinfo {volume} {89}},\ \bibinfo
  {pages} {1} (\bibinfo {year} {2016})}\BibitemShut {NoStop}%
\bibitem [{\citenamefont {Bosse}\ \emph {et~al.}(2011)\citenamefont {Bosse},
  \citenamefont {Pathak},\ and\ \citenamefont {Singh}}]{Bosse2011}%
  \BibitemOpen
  \bibfield  {author} {\bibinfo {author} {\bibfnamefont {J.}~\bibnamefont
  {Bosse}}, \bibinfo {author} {\bibfnamefont {K.~N.}\ \bibnamefont {Pathak}},\
  and\ \bibinfo {author} {\bibfnamefont {G.~S.}\ \bibnamefont {Singh}},\
  }\href@noop {} {\bibfield  {journal} {\bibinfo  {journal} {Phys. Rev. E}\
  }\textbf {\bibinfo {volume} {84}},\ \bibinfo {pages} {042101} (\bibinfo
  {year} {2011})}\BibitemShut {NoStop}%
\bibitem [{\citenamefont {Dornheim}\ \emph
  {et~al.}(2016{\natexlab{b}})\citenamefont {Dornheim}, \citenamefont {Groth},
  \citenamefont {Schoof}, \citenamefont {Hann},\ and\ \citenamefont
  {Bonitz}}]{Dornheim2016b}%
  \BibitemOpen
  \bibfield  {author} {\bibinfo {author} {\bibfnamefont {T.}~\bibnamefont
  {Dornheim}}, \bibinfo {author} {\bibfnamefont {S.}~\bibnamefont {Groth}},
  \bibinfo {author} {\bibfnamefont {T.}~\bibnamefont {Schoof}}, \bibinfo
  {author} {\bibfnamefont {C.}~\bibnamefont {Hann}},\ and\ \bibinfo {author}
  {\bibfnamefont {M.}~\bibnamefont {Bonitz}},\ }\bibfield  {title} {\bibinfo
  {title} {Ab initio quantum monte carlo simulations of the uniform electron
  gas without fixed nodes: The unpolarized case},\ }\href@noop {} {\bibfield
  {journal} {\bibinfo  {journal} {Phys. Rev. B}\ }\textbf {\bibinfo {volume}
  {93}},\ \bibinfo {pages} {205134} (\bibinfo {year}
  {2016}{\natexlab{b}})}\BibitemShut {NoStop}%
\bibitem [{\citenamefont {Groth}\ \emph {et~al.}(2016)\citenamefont {Groth},
  \citenamefont {Schoof}, \citenamefont {Dornheim},\ and\ \citenamefont
  {Bonitz}}]{Groth2016}%
  \BibitemOpen
  \bibfield  {author} {\bibinfo {author} {\bibfnamefont {S.}~\bibnamefont
  {Groth}}, \bibinfo {author} {\bibfnamefont {T.}~\bibnamefont {Schoof}},
  \bibinfo {author} {\bibfnamefont {T.}~\bibnamefont {Dornheim}},\ and\
  \bibinfo {author} {\bibfnamefont {M.}~\bibnamefont {Bonitz}},\ }\bibfield
  {title} {\bibinfo {title} {Ab initio quantum monte carlo simulations of the
  uniform electron gas without fixed nodes},\ }\href@noop {} {\bibfield
  {journal} {\bibinfo  {journal} {Phys. Rev. B}\ }\textbf {\bibinfo {volume}
  {93}},\ \bibinfo {pages} {085102} (\bibinfo {year} {2016})}\BibitemShut
  {NoStop}%
\bibitem [{\citenamefont {Melik-Alaverdian}\ \emph {et~al.}(2001)\citenamefont
  {Melik-Alaverdian}, \citenamefont {Ortiz},\ and\ \citenamefont
  {Bonesteel}}]{Melik2001}%
  \BibitemOpen
  \bibfield  {author} {\bibinfo {author} {\bibfnamefont {V.}~\bibnamefont
  {Melik-Alaverdian}}, \bibinfo {author} {\bibfnamefont {G.}~\bibnamefont
  {Ortiz}},\ and\ \bibinfo {author} {\bibfnamefont {N.~E.}\ \bibnamefont
  {Bonesteel}},\ }\bibfield  {title} {\bibinfo {title} {Quantum projector
  method on curved manifolds},\ }\href@noop {} {\bibfield  {journal} {\bibinfo
  {journal} {J. Stat. Phys.}\ }\textbf {\bibinfo {volume} {104}},\ \bibinfo
  {pages} {449} (\bibinfo {year} {2001})}\BibitemShut {NoStop}%
\bibitem [{\citenamefont {Fantoni}(2018{\natexlab{b}})}]{Fantoni2018}%
  \BibitemOpen
  \bibfield  {author} {\bibinfo {author} {\bibfnamefont {R.}~\bibnamefont
  {Fantoni}},\ }\href@noop {} {\bibfield  {journal} {\bibinfo  {journal} {Int.
  J. Mod. Phys. C}\ }\textbf {\bibinfo {volume} {29}},\ \bibinfo {pages}
  {1850064} (\bibinfo {year} {2018}{\natexlab{b}})}\BibitemShut {NoStop}%
\bibitem [{\citenamefont {Trotter}(1959)}]{Trotter1959}%
  \BibitemOpen
  \bibfield  {author} {\bibinfo {author} {\bibfnamefont {H.~F.}\ \bibnamefont
  {Trotter}},\ }\href@noop {} {\bibfield  {journal} {\bibinfo  {journal} {Proc.
  Am. Math. Soc.}\ }\textbf {\bibinfo {volume} {10}},\ \bibinfo {pages} {545}
  (\bibinfo {year} {1959})}\BibitemShut {NoStop}%
\bibitem [{\citenamefont {Simon}(1979)}]{Simon1979}%
  \BibitemOpen
  \bibfield  {author} {\bibinfo {author} {\bibfnamefont {B.}~\bibnamefont
  {Simon}},\ }\href@noop {} {\emph {\bibinfo {title} {Functional integration
  and quantum physics}}}\ (\bibinfo  {publisher} {Academic},\ \bibinfo
  {address} {New York},\ \bibinfo {year} {1979})\BibitemShut {NoStop}%
\bibitem [{\citenamefont {Abramowitz}\ and\ \citenamefont
  {Stegun}(1970)}]{Abramowitz}%
  \BibitemOpen
  \bibfield  {author} {\bibinfo {author} {\bibfnamefont {M.}~\bibnamefont
  {Abramowitz}}\ and\ \bibinfo {author} {\bibfnamefont {I.~A.}\ \bibnamefont
  {Stegun}},\ }\href@noop {} {\emph {\bibinfo {title} {Handbook of mathematical
  functions}}}\ (\bibinfo  {publisher} {Dover},\ \bibinfo {address} {New
  York},\ \bibinfo {year} {1970})\BibitemShut {NoStop}%
\bibitem [{\citenamefont {Barker}(1979)}]{Barker1979}%
  \BibitemOpen
  \bibfield  {author} {\bibinfo {author} {\bibfnamefont {J.~A.}\ \bibnamefont
  {Barker}},\ }\href@noop {} {\bibfield  {journal} {\bibinfo  {journal} {J.
  Chem. Phys.}\ }\textbf {\bibinfo {volume} {70}},\ \bibinfo {pages} {2914}
  (\bibinfo {year} {1979})}\BibitemShut {NoStop}%
\end{thebibliography}%

\end{document}